\documentclass[journal=jctcce, manuscript=article]{achemso}

\usepackage[utf8]{inputenc}
\usepackage[T1]{fontenc}
\usepackage{amsmath}
\usepackage{amssymb}
\usepackage{graphicx}
\usepackage{xcolor}
\usepackage{booktabs}
\usepackage{multirow, multicol}
\usepackage{geometry}
\usepackage{siunitx}
\DeclareSIUnit\hartree{\text {\ensuremath {E}}_{\mathrm {h}}}
\DeclareSIUnit\angstrom{\text {Å}}
\usepackage{array}
\usepackage{threeparttable}
\usepackage[version=4]{mhchem}
\usepackage{hyperref}
\usepackage{braket}
\hypersetup{colorlinks=true, linkcolor=blue, citecolor=blue, urlcolor=blue}
\usepackage{tikz}
\usetikzlibrary{shapes,arrows,positioning,calc,fit,backgrounds,shapes.geometric,arrows.meta}

\SectionNumbersOn

\author{Joshua J. Goings}
\affiliation{
IonQ Inc, College Park, MD, 20740, USA
}
\author{Kyujin Shin}
\email{shinkj@hyundai.com}
\affiliation{
 Materials Research \& Engineering Center, Advanced Vehicle Platform Division, Hyundai Motor Company, Uiwang 16082, Republic of Korea
}

\author{Seunghyo Noh}
\affiliation{ 
  Materials Research \& Engineering Center, Advanced Vehicle Platform Division, Hyundai Motor Company, Uiwang 16082, Republic of Korea
}

\author{Woomin Kyoung}
\affiliation{ 
  Materials Research \& Engineering Center, Advanced Vehicle Platform Division, Hyundai Motor Company, Uiwang 16082, Republic of Korea
}

\author{Donghwi Kim}
\affiliation{ 
  Materials Research \& Engineering Center, Advanced Vehicle Platform Division, Hyundai Motor Company, Uiwang 16082, Republic of Korea
}

\author{Jihye Baek}
\affiliation{ 
  Materials Research \& Engineering Center, Advanced Vehicle Platform Division, Hyundai Motor Company, Uiwang 16082, Republic of Korea
}

\author{Martin Roetteler}
\affiliation{
IonQ Inc, College Park, MD, 20740, USA
}

\author{Evgeny Epifanovsky}
\affiliation{
IonQ Inc, College Park, MD, 20740, USA
}

\author{Luning Zhao}
\email{zhao@ionq.co}
\affiliation{
IonQ Inc, College Park, MD, 20740, USA
}
\title{Molecular Properties in Quantum-Classical Auxiliary-Field Quantum Monte Carlo: Correlated Sampling with Application to Accurate Nuclear Forces} 

\begin{document}

\begin{tocentry}
\begin{center}
\includegraphics[height=4.5cm]{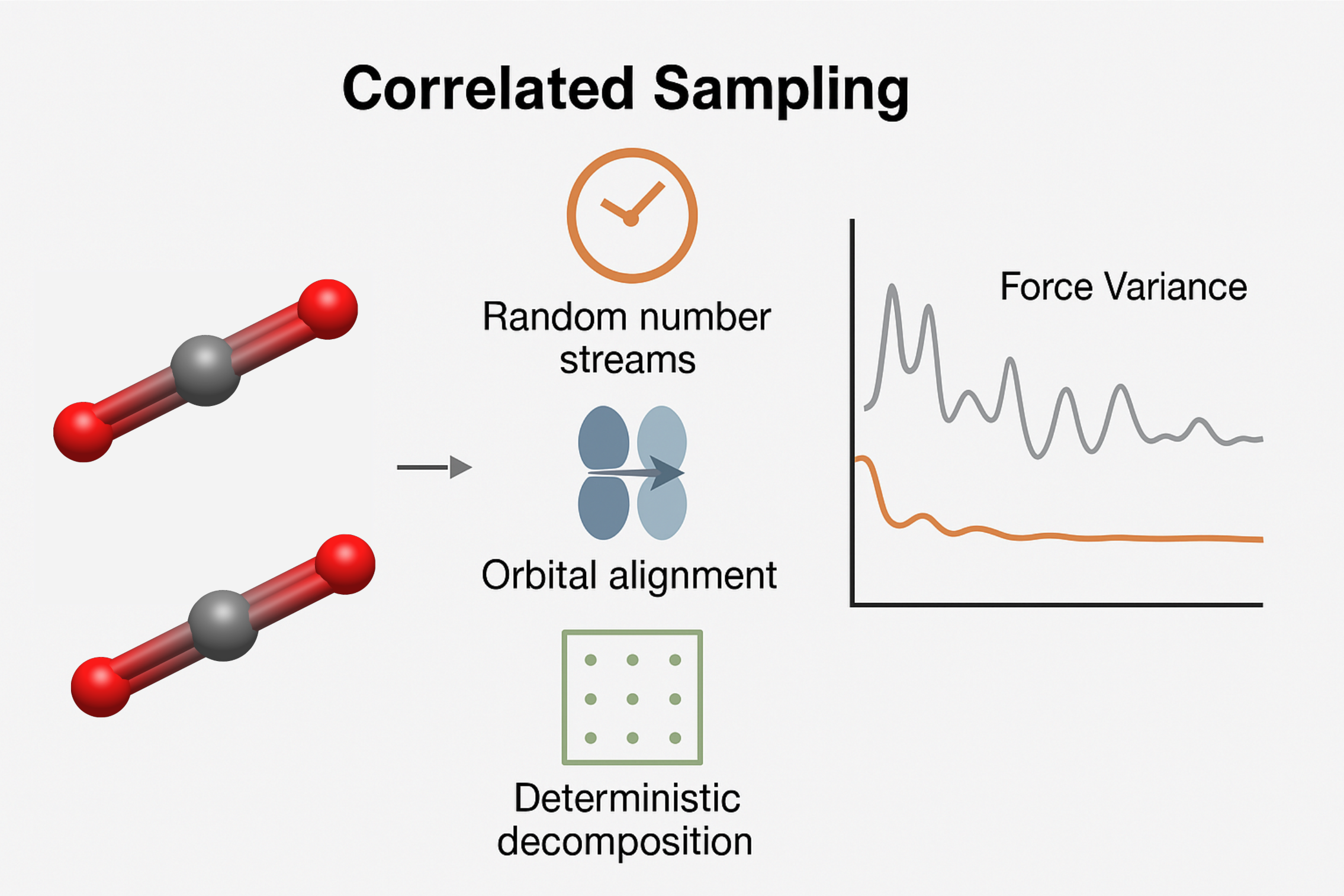} 

\end{center}
Accurate nuclear forces for strongly correlated molecules are computed using quantum-classical auxiliary-field quantum Monte Carlo (QC-AFQMC) combined with a robust correlated sampling technique. This enables geometry optimizations and reaction studies on near-term quantum devices by drastically reducing statistical noise in energy differences.
\end{tocentry}

\begin{abstract}
We extend correlated sampling from classical auxiliary-field quantum Monte Carlo to the quantum-classical (QC-AFQMC) framework, enabling accurate nuclear force computations crucial for geometry optimization and reaction dynamics. Stochastic electronic structure methods typically encounter prohibitive statistical noise when computing gradients via finite differences. To address this, our approach maximizes correlation between nearby geometries by synchronizing random number streams, aligning orbitals, using deterministic integral decompositions, and employing a consistent set of classical shadow measurements defined at a single reference geometry. Crucially, reusing this single, reference-defined shadow ensemble eliminates the need for additional quantum measurements at displaced geometries. Together, these methodological choices substantially reduce statistical variance in computed forces. We validate the method across hydrogen chains, confirming accuracy throughout varying correlation regimes, and demonstrate significant improvements over single-reference methods in force evaluations for N$_2$ and stretched linear H$_4$, particularly in strongly correlated regions where conventional coupled cluster approaches qualitatively fail. Orbital-optimized trial wavefunctions further boost accuracy for demanding cases such as stretched CO$_2$, without increasing quantum resource requirements. Finally, we apply our methodology to the MEA-CO$_2$ carbon capture reaction, employing quantum information metrics for active space selection and matchgate shadows for efficient overlap evaluations, establishing QC-AFQMC as a robust framework for exploring complex reaction pathways.
\end{abstract}
\section{Introduction}

A central challenge in electronic structure theory is the accurate description of electron correlation. Correlation effects span a continuum, but are often conceptually divided into two regimes. Dynamic correlation involves the cumulative effect of numerous high-energy, small-amplitude configurations, often well-captured by methods like coupled cluster theory.\cite{Bartlett2007-kj} Static (or strong) correlation, conversely, arises when two or more electronic configurations contribute with comparable weight to the ground state wavefunction, a situation prevalent in bond breaking, transition metal chemistry, and excited states.\cite{Park2020-nw} This dichotomy presents a significant hurdle: efficient methods adept at dynamic correlation often fail catastrophically in the presence of strong static correlation, while approaches designed for static correlation, such as multiconfigurational methods\cite{Roos2016-gs} or density matrix renormalization group theory,\cite{Chan2011-ad} are typically prohibitively expensive to apply, and as a result require restricting the correlation treatment to a predefined active space, potentially neglecting crucial dynamic correlation effects outside this subspace. Overcoming these limitations is essential for quantitatively modeling many chemically significant processes.

Stochastic methods, particularly quantum Monte Carlo (QMC) techniques \cite{Lee2022-rl, Motta2018-ft}, offer a compelling alternative route towards achieving high accuracy across the correlation spectrum, often exhibiting favorable polynomial scaling with system size. Among these, auxiliary-field QMC (AFQMC) has emerged as a particularly promising approach.\cite{Zhang2018-dn,Lee2022-rl, Shee2023-gs} Furthermore, recent developments integrating quantum computation have led to quantum-enhanced AFQMC (QC-AFQMC) \cite{Huggins2022-lh, Wan2023-df, Kiser2024-be, Huang2024-gz, Zhao2025-fb}, which aims to achieve accuracy approaching that of full configuration interaction within a given basis set. The advantage of QC-AFQMC lies in efficiently evaluating overlaps between complex trial wavefunctions and the Monte Carlo walkers using quantum circuits \cite{Lee2022-rl, Jiang2025-gc, Huang2024-gz, Zhao2025-fb}. This allows for more expressive trial states that better constrain the challenging phase problem inherent in AFQMC, potentially bridging the gap between accuracy and computational feasibility for strongly correlated systems.

While QMC methods provide powerful pathways to accurate ground- or excited-state energies, a practical challenge arises when computing properties defined by energy derivatives with respect to parameters such as nuclear coordinates (forces), external fields (polarizabilities), or particle number (chemical potentials). Unlike deterministic methods such as density functional theory or coupled cluster theory, which have greatly benefited from the development of analytic energy gradient techniques \cite{Yamaguchi1994-pq,Handy1984-yf,Salter1989-sw,Salter1989-oa,Helgaker1988-qx,Rice1985-ea,Pople2009-qq,Neese2009-rr,Park2020-nw,Gauss1991-ss,Bakken1999-sw}, the inherent statistical uncertainty in stochastic energy evaluations complicates direct differentiation. Specialized techniques are therefore required to manage statistical noise effectively when computing energy differences \cite{Assaraf2000-jh, Poole2014-vr}.

Correlated sampling provides a mathematically general and statistically powerful framework for precisely extracting such energy differences from stochastic calculations \cite{Shee2017-xp, Chen2023-pz}. By carefully engineering strong positive correlations between stochastic estimates performed at slightly perturbed parameter values (e.g., infinitesimally displaced geometries), this technique dramatically reduces the statistical variance in the energy \textit{difference}, often by orders of magnitude, even when the absolute energies themselves retain significant statistical uncertainty \cite{Shee2017-xp, Chen2023-pz}. This variance reduction is crucial for reliably computing energy gradients. While numerical differentiation via correlated sampling involves different computational considerations compared to purely analytic gradients, its generality unlocks access to a broad spectrum of chemically and physically important properties within high-accuracy stochastic methods. It is complementary to other techniques like backpropagation within AFQMC \cite{Motta2017-zg, Mahajan2023-vf}, which directly targets expectation values rather than energy differences but also enables property calculations. Correlated sampling, however, offers a distinct advantage for computing energy derivatives with complex trial states. Extending backpropagation through the full intricacies of the AFQMC algorithm---including the stochastic propagation and potentially complex trial function evaluation---can present significant implementation hurdles, which correlated sampling naturally bypasses by focusing on correlated differences between separate simulations.

Classical implementations of AFQMC have already established correlated sampling as a potent tool for property evaluation \cite{Motta2018-hr, Chen2023-yv, Mahajan2023-vf, Shee2017-xp}. However, extending this technique to the hybrid quantum-classical framework of QC-AFQMC introduces unique challenges at the interface between the deterministic quantum circuit evaluations and the stochastic classical propagation. Maintaining the strong statistical correlation necessary for variance reduction requires careful orchestration across this boundary. In this work, we adapt and implement correlated sampling techniques specifically for QC-AFQMC, focusing on four critical synchronization and consistency mechanisms: (1) the use of consistent random number sequences for the evolution of corresponding walkers in perturbed and unperturbed simulations, (2) rigorous orbital alignment procedures to preserve the physical character of the trial wavefunction across perturbations, (3) the use of deterministic two-electron integral decomposition strategies to maintain consistent auxiliary-field representations, and (4) the consistent application of classical shadow measurement ensembles derived from the reference geometry. The coordinated application of these elements maximizes statistical correlation between energy evaluations, substantially reducing uncertainty in computed property differences.

We focus on nuclear forces as a particularly demanding and chemically vital test case for this methodology. Accurate forces are essential for geometry optimizations, transition state searches, and molecular dynamics simulations \cite{Shee2017-xp, Chen2023-pz}. They require high precision in energy differences and serve as sensitive probes of imperfect correlation between stochastic runs. We demonstrate the capability of QC-AFQMC with correlated sampling for force calculations on systems like hydrogen chains (H$_n$) and N$_2$ dissociation, which are chosen to exhibit significant static correlation where methods like coupled cluster can fail qualitatively. Furthermore, for challenging, strongly correlated cases such as stretched CO$_2$, we investigate the impact of trial wavefunction quality on overall accuracy, showing that employing orbital-optimized unitary pair coupled cluster doubles (upCCD) trial wavefunctions significantly enhances the description of the system without increasing quantum resource requirements compared to standard upCCD.

To illustrate the practical relevance and establish a complete workflow, we apply the methodology to compute interaction energies and forces relevant to the reaction between monoethanolamine (MEA) and CO$_2$, a key process in industrial carbon capture technology. This application integrates the correlated sampling force evaluation with techniques like quantum information metrics for active space selection \cite{Stein2016-yg} and matchgate shadows for efficient quantum overlap estimation \cite{Wan2023-df, Huang2024-gz, Zhao2025-fb}.

While nuclear forces serve as a rigorous validation metric, the correlated sampling framework developed here is readily extensible to other molecular properties determined by energy differences, such as ionization potentials, electron affinities, proton affinities, and reaction energy barriers.\cite{Shee2017-xp} By integrating the potential advantages of quantum computation with robust statistical variance reduction techniques, this work offers a powerful paradigm for expanding the domain of applicability for QMC methods, enabling reliable prediction of chemical properties even in the challenging realm of strongly correlated molecular systems.

\section{Theory}

QMC methods provide a powerful stochastic approach for achieving high accuracy in electronic structure calculations. By employing stochastic sampling, they offer a path towards high accuracy for complex systems, often exhibiting favorable computational scaling --- potentially polynomial under certain approximations --- but at the price of introducing statistical uncertainty. The AFQMC approach,\cite{Umrigar2015-zv,Zhang2018-dn,Motta2018-ft,Motta2019-ki,Lee2020-gp,Lee2020-sm,Shi2021-ke,Mahajan2021-ly,Mahajan2022-aa,Lee2022-rl,Malone2023-kl,Shee2023-gs,Pham2024-nq,Sukurma2024-cg,Jiang2024-rp,Jiang2025-gc} in particular, has emerged as a powerful framework because it transforms the many-body problem into a statistical integration over auxiliary fields coupled to one-body operators. This auxiliary-field decomposition is highly advantageous. It recasts the interacting problem in a form amenable to stochastic sampling using efficient polynomial-scaling operations on independent-particle propagators (like Slater determinants). Equally importantly, this framework provides a practical means, typically involving importance sampling guided by a trial wavefunction, to control the severe fermion sign (or phase) problem that often plagues fermionic QMC methods, thus enabling stable routes to high accuracy. 

\subsection{Quantum-Classical AFQMC Framework}

AFQMC achieves ground state properties through stochastic evolution in imaginary time. The core mathematical insight lies in representing quantum states as linear combinations of Slater determinants, or ``walkers,'' whose evolution follows from a Hamiltonian decomposed into manageable one- and two-body terms. This evolution assumes a particularly elegant form through the Hubbard-Stratonovich transformation,\cite{Stratonovich1957-hn,Hubbard1959-ma} which recasts two-body interactions as an integration over auxiliary fields that couple to one-body operators. By sampling these auxiliary fields from their normal distribution, we transform a quantum mechanical problem into a stochastic process with a well-defined stationary distribution.

The fundamental equation governing this process expresses the ground state as the asymptotic limit of imaginary time evolution:

\begin{equation}
    \ket{\Psi_0} = \lim_{\tau \to \infty} e^{-\tau H} \ket{\Phi_0} = \lim_{\tau \to \infty} \ket{\Psi(\tau)}
\end{equation}

This equation captures how the ground state $\ket{\Psi_0}$ emerges from propagating an initial state $\ket{\Phi_0}$ through imaginary time. At each step in this evolution, the wave function maintains its representation as a weighted superposition of walkers:

\begin{equation}
    \ket{\Psi(\tau)} = \sum_i w_i(\tau) \ket{\phi_i(\tau)}
\end{equation}

Here, $w_i(\tau)$ represents the statistical weight of each walker $\ket{\phi_i(\tau)}$ at imaginary time $\tau$.

The direct application of this method, however, encounters the infamous ``sign problem'' or, more generally, the ``phase problem,'' where walkers acquire complex phases that cannot be interpreted as probabilities. To circumvent this fundamental difficulty, AFQMC employs a trial wave function to guide the simulation, constraining the paths to maintain a positive real part of the overlap with this reference state. Each walker's weight evolves throughout the simulation to more accurately represent the projected ground state properties.

The electronic Hamiltonian, which drives this evolution, decomposes into one- and two-body terms:

\begin{equation}
    H = H_1 + H_2 = \sum_{ij}^N T_{ij} c^\dagger_i c_j + \frac{1}{2} \sum_{ijkl}^N V_{ijkl} c^\dagger_i c^\dagger_j c_k c_l
\end{equation}

The crucial mathematical transformation occurs with the two-body term, which can be expressed as a sum of squares of one-body operators:

\begin{equation}
    H_2 = -\frac{1}{2} \sum_\alpha \lambda_\alpha v_\alpha^2
\end{equation}

Through the Hubbard-Stratonovich transformation, this structure allows us to convert the exponential of $H_2$ into an integral over auxiliary fields:

\begin{equation}
    e^{-\tau H_2} = \prod_\alpha \frac{1}{\sqrt{2\pi}} \int_{-\infty}^{\infty} e^{-\frac{1}{2}x_\alpha^2} e^{\sqrt{\tau} x_\alpha \sqrt{\lambda_\alpha} v_\alpha} dx_\alpha
\end{equation}

This expression demonstrates how the many-body propagator transforms into a statistical integral over auxiliary fields $x_\alpha$, each coupled to a one-body operator. The integration can be performed stochastically by sampling the auxiliary fields from their normal distribution, resulting in a collection of one-body propagators acting on Slater determinants.

Energy evaluation within this framework employs a mixed estimator that combines the trial state with the stochastically evolved state:

\begin{equation}
    E(\tau) = \frac{\braket{\Psi_T | H | \Psi(\tau)}}{\braket{\Psi_T | \Psi(\tau)}} = \frac{\sum_i w_i E^{(i)}(\tau)}{\sum_i w_i}
\label{eq:mixed_estimator}
\end{equation}

where $E^{(i)}(\tau)$ represents the local energy of each walker:

\begin{equation}
    E^{(i)}(\tau) = \frac{\braket{\Psi_T | H | \phi_i(\tau)}}{\braket{\Psi_T | \phi_i(\tau)}}
\end{equation}

The accuracy and efficiency of AFQMC calculations critically depend on the quality of the trial wave function $|\Psi_T\rangle$. Classical implementations typically employ the Hartree-Fock state due to its computational simplicity when evaluating overlaps. However, this choice becomes inadequate for systems with significant multiconfigurational character. More expressive ans{\"a}tze derived from coupled cluster methods would offer greater accuracy, but they traditionally incur exponential computational costs when evaluating overlaps with arbitrary Slater determinants.

Quantum-enhanced AFQMC (QC-AFQMC) resolves this fundamental tension.\cite{Huggins2022-lh,Xu2022-xs,Wan2023-df,Amsler2023-hr,Kiser2024-be,Huang2024-gz,Kiser2025-ot} By leveraging quantum computing resources, it enables efficient measurement of overlaps with complex trial states, removing the exponential bottleneck associated with correlated wave functions. The quantum advantage operates precisely at the interface between the trial state and the stochastic evolution, that is, the overlaps between the quantum-prepared trial state and the classically evolved walkers.

The central quantum-computed value involves measuring:

\begin{equation}
    \bra{\Psi_T} H \ket{\phi(\tau)} = \sum_{pr} \braket{\Psi_T | \phi_p^r} \bra{\phi_p^r} H \ket{\phi(\tau)} + \sum_{pqrs} \braket{\Psi_T | \phi_{pq}^{rs}} \bra{\phi_{pq}^{rs}} H \ket{\phi(\tau)}
\end{equation}

This expression decomposes the Hamiltonian element into contributions from single and double excitations from the walker state, each requiring overlap measurements between the trial state and the excited determinants.

While the Hadamard test provides a direct approach to measuring these overlaps, it requires a separate quantum circuit execution for each walker and each excitation, potentially leading to prohibitive quantum resource requirements for systems with many walkers. Furthermore, the Hadamard test necessitates controlled operations that significantly increase circuit depth and gate count, making it impractical for near-term quantum devices with limited coherence times and gate fidelities.\cite{Nielsen2010-dm, Huggins2022-lh}

The matchgate shadows protocol offers a more efficient alternative by exploiting the symmetry structure of fermionic systems.\cite{Wan2023-df} This approach samples from an ensemble of unitary transformations that preserve fermionic exchange statistics, measures the transformed states in the computational basis, and reconstructs the desired overlaps through classical post-processing. Recent implementations have demonstrated the practical scalability of this approach for quantum chemistry applications.\cite{Zhao2025-fb} The approach is based on the more general concept of ``classical shadows''.\cite{Huang2020-xp, Huang2022-wy} Crucially, matchgate circuits naturally encode fermionic operations under the Jordan-Wigner mapping, requiring only nearest-neighbor interactions that are more amenable to implementation on current quantum architectures. Furthermore, their structure allows for efficient classical simulation of certain fermionic systems, providing a valuable benchmark for quantum advantage.

The overlap estimate for each walker takes the form:

\begin{equation}
\langle\Psi_T|\phi_w\rangle \approx \frac{1}{N_s}\sum_{i=1}^{N_s} f(b_i, U_i, \phi_w)
\label{eq:shadow_estimator}
\end{equation}

where $N_s$ represents the number of shadow samples, $b_i$ the measurement outcomes, $U_i$ the applied matchgate circuits, and $f$ a reconstruction function specific to matchgate shadows. Critically, this approach allows simultaneous estimation of overlaps for multiple walkers from the same set of measurements, dramatically reducing the quantum resource requirements compared to direct Hadamard tests.

The matchgate shadow approach achieves polynomial scaling with system size:

\begin{equation}
    \mathcal{O}\left(\log(\frac{M}{\delta}) \frac{b_{\text{max}}}{\epsilon^2}\right)
\end{equation}

where $M$ represents the number of walkers, $\delta$ a probability parameter, $\epsilon$ the desired error in the overlap, and $b_{\text{max}}$ scales as $O(\sqrt{n} \log n)$ with $n$ being the number of qubits.\cite{Wan2023-df} This polynomial scaling represents a dramatic improvement over the exponential costs typically associated with quantum simulation of strongly correlated systems.

\subsection{Forces via Correlated Sampling}

Computing nuclear forces (the derivatives of electronic energy with respect to nuclear coordinates) presents a distinctive challenge for stochastic methods like QC-AFQMC. While the Hellmann-Feynman theorem\cite{Feynman1939-sl} provides an elegant formalism for analytical derivatives in deterministic approaches, its application to stochastic wavefunctions introduces fundamental complications related to statistical uncertainties and the need for Pulay terms to account for basis set effects.\cite{Pulay1969-dj}

In practice, most QMC implementations employ finite difference approximations for force components:
\begin{equation}
F_i = -\frac{\partial E}{\partial R_i} \approx -\frac{E(\mathbf{R} + \delta\mathbf{e}_i) - E(\mathbf{R} - \delta\mathbf{e}_i)}{2\delta}
\label{eq:finite_difference}
\end{equation}
where $\mathbf{R}$ represents nuclear coordinates and $\delta$ a small displacement along direction $\mathbf{e}_i$. This approach introduces a formidable challenge: each energy evaluation $E(\mathbf{R} \pm \delta\mathbf{e}_i)$ carries statistical uncertainty $\sigma_E$, which propagates into the force estimates.

Simple error propagation reveals the variance in force components:
\begin{equation}
\sigma^2_{F_i} = \frac{1}{4\delta^2}[\sigma^2_{E_+} + \sigma^2_{E_-} - 2\mathrm{Cov}(E_+, E_-)]
\label{eq:force_variance_raw}
\end{equation}
where $E_+$ and $E_-$ represent energies at $\mathbf{R} + \delta\mathbf{e}_i$ and $\mathbf{R} - \delta\mathbf{e}_i$, respectively, and $\mathrm{Cov}(E_+, E_-)$ denotes their covariance. By defining the correlation coefficient $\rho$ between these energy evaluations:
\begin{equation}
\rho = \frac{\mathrm{Cov}(E_+, E_-)}{\sigma_{E_+}\sigma_{E_-}}
\label{eq:correlation_coeff}
\end{equation}
and assuming approximate equality of energy variances ($\sigma_{E_+} \approx \sigma_{E_-} \approx \sigma_E$), the force variance simplifies to:
\begin{equation}
\sigma^2_{F_i} \approx \frac{\sigma^2_E (1-\rho)}{2\delta^2}
\label{eq:force_variance_correlated}
\end{equation}

This expression reveals the central insight of correlated sampling: as the correlation coefficient $\rho$ approaches unity, the force variance can be dramatically reduced even if the absolute energy uncertainties $\sigma_E$ remain significant. Thus, the primary challenge in QC-AFQMC force calculations lies in maximizing correlation between energy evaluations at slightly displaced geometries.

\subsubsection{Variance Reduction through Correlation}

To illustrate why correlation so effectively reduces variance, consider a simplified model where energy evaluations at displaced geometries can be expressed as:
\begin{equation}
E_+ = E_{\text{true}} + \frac{\Delta E}{2} + \epsilon_+, \quad E_- = E_{\text{true}} - \frac{\Delta E}{2} + \epsilon_-
\end{equation}
where $\Delta E = E(\mathbf{R} + \delta\mathbf{e}_i) - E(\mathbf{R} - \delta\mathbf{e}_i)$ represents the true energy difference and $\epsilon_+$, $\epsilon_-$ are statistical fluctuations (with mean zero and variance $\sigma^2_E$).

Without correlation ($\epsilon_+$ and $\epsilon_-$ independent), the force estimator has variance:
\begin{equation}
\text{Var}(F_i) = \text{Var}\left(-\frac{E_+ - E_-}{2\delta}\right) = \text{Var}\left(-\frac{\Delta E + (\epsilon_+ - \epsilon_-)}{2\delta}\right) = \frac{\sigma^2_E}{2\delta^2}
\end{equation}

With perfect correlation ($\epsilon_+ = \epsilon_-$), the fluctuations cancel exactly:
\begin{equation}
F_i = -\frac{E_+ - E_-}{2\delta} = -\frac{\Delta E + (\epsilon_+ - \epsilon_-)}{2\delta} = -\frac{\Delta E}{2\delta}
\end{equation}
yielding the exact force (within the finite difference approximation) with zero statistical variance.

\subsubsection{Correlated Sampling Implementation}

Implementing effective correlated sampling in QC-AFQMC requires systematic control of stochastic elements at multiple levels. Our hierarchical strategy addresses four primary sources of randomness:

\begin{enumerate}
\item {\textit{Random Number Stream Control.}
The foundation of correlated sampling is synchronizing random number generation across calculations at different geometries. By seeding pseudo-random number generators identically for corresponding calculations at $\mathbf{R} + \delta\mathbf{e}_i$ and $\mathbf{R} - \delta\mathbf{e}_i$ (and potentially the reference geometry $\mathbf{R}$ if using forward/backward difference), we ensure that Monte Carlo walkers encounter identical sequences of auxiliary fields during propagation, creating strong statistical correlations between energy evaluations.}

\item {\textit{Orbital Alignment Protocol.}
Nuclear displacements modify the molecular orbital basis, potentially altering orbital character or ordering even for small geometric perturbations. Since QC-AFQMC calculations depend critically on the one-particle basis, maintaining consistent orbital representations across geometries is essential for effective correlation.

We implement a rigorous orbital alignment procedure based on maximizing overlap between orbitals at a reference geometry and the target (displaced) geometry. First, we obtain molecular orbital coefficient matrices $C_{\text{ref}}$ and $C_{\text{target}}$ and construct orthonormalized orbitals:
\begin{equation}
\tilde{C}_{\text{ref}} = S_{\text{ref}}^{-1/2} C_{\text{ref}}, \quad \tilde{C}_{\text{target}} = S_{\text{target}}^{-1/2} C_{\text{target}}
\label{eq:orthonormalize_orbitals}
\end{equation}
where $S_{\text{ref}}$ and $S_{\text{target}}$ are the corresponding atomic orbital overlap matrices.

Special attention is required for nearly degenerate orbitals, defined as those with energy differences below a threshold $\delta_{\text{thresh}}$:
\begin{equation}
G_k = \{(i,j) : |\epsilon_i - \epsilon_j| < \delta_{\text{thresh}}\}
\end{equation}

For each degenerate subspace $G_k$, we compute the overlap matrix between reference and target orbitals within that subspace:
\begin{equation}
O_k = \tilde{C}^{\dagger}_{\text{ref},k} \tilde{C}_{\text{target},k}
\label{eq:subspace_overlap}
\end{equation}
where the matrices are restricted to the columns corresponding to subspace $k$. If the atomic orbital bases differ between geometries, an additional transformation involving $S_{\text{inter}}$ (the overlap between the two AO bases) would be necessary here.

We perform a singular value decomposition (SVD) on this overlap matrix:
\begin{equation}
O_k = U_k\Sigma_kV^{\dagger}_k
\label{eq:svd_overlap}
\end{equation}

The optimal unitary rotation matrix that maximizes overlap within this subspace is:
\begin{equation}
R_k = U_kV^{\dagger}_k
\label{eq:rotation_matrix}
\end{equation}

For real-valued orbitals, we apply phase correction if $\det(R_k) < 0$, replacing $R_k$ with $R_kP$ where $P$ is a diagonal matrix with elements $(1,...,1,-1)$.

The target orbitals are then rotated within each subspace $k$ using $R_k$. For non-degenerate orbitals, the alignment is typically trivial (identity rotation) unless reordering occurred. The final aligned target orbitals $C_{\text{aligned}}$ are constructed by applying these rotations appropriately across all orbital subspaces.}

\item {\textit{Deterministic Two-Electron Integral Decomposition.}
The final component of our correlated sampling strategy addresses the decomposition of two-electron integrals, which determine the structure of auxiliary fields. In AFQMC, four-index integrals $(pq|rs)$ are expressed through a modified Cholesky decomposition:\cite{Motta2019-ki}
\begin{equation}
V_{(pq)(rs)} = (pq|rs) = \sum_\gamma L^{\gamma}_{pq}L^{\gamma}_{rs}
\label{eq:cholesky_decomp}
\end{equation}
where indices $pq$ and $rs$ are treated as composite row and column indices in a positive definite matrix representation of the two-electron operator.

The standard Cholesky algorithm proceeds iteratively. At each step $k$, it computes the residual diagonal elements:
\begin{equation}
D_{pq} = V_{(pq)(pq)} - \sum_{\gamma=1}^{k-1}(L^{\gamma}_{pq})^2
\label{eq:cholesky_diag}
\end{equation}

A pivot index pair $(\bar{p},\bar{q})$ is selected, typically by maximizing the diagonal element $D_{pq}$, and the corresponding Cholesky vector is computed:
\begin{equation}
L^{(k)}_{pq} = \frac{V_{(pq)(\bar{p}\bar{q})} - \sum_{\gamma=1}^{k-1}L^{\gamma}_{pq}L^{\gamma}_{\bar{p}\bar{q}}}{\sqrt{D_{\bar{p}\bar{q}}}}
\label{eq:cholesky_vector}
\end{equation}

While mathematically equivalent decompositions can arise from different pivot selection sequences, they yield distinct representations of the auxiliary fields $\hat{B}(\mathbf{x})$ in Eq.~\ref{eq:cholesky_decomp}. To ensure correlation, we fix the pivot selection sequence determined at the reference geometry ($\mathbf{R}$) and enforce the use of this same sequence when decomposing the integrals at the displaced geometries ($\mathbf{R} \pm \delta\mathbf{e}_i$). This ensures the auxiliary-field representation remains consistent throughout the force calculation.}
\item {\textit{Consistent Classical Shadow Measurements.}
When estimating observables via classical shadows, an essential aspect of correlated sampling is consistency in the measurement ensemble. We define the set of random matchgate unitaries used for shadow measurements solely from the trial state at the reference geometry $\mathbf{R}$, eliminating additional quantum shot budgets and ensuring that quantum measurement resources remain confined to the reference. By reusing exactly this ensemble for each perturbed geometry $\mathbf{R}\pm\delta\mathbf{e}_i$, we significantly reduce computational overhead compared to generating independent shadows at each geometry. Crucially, this consistency also locks in statistical noise correlations arising from finite shadow sampling, which dramatically reduces variance in computed energy differences.}
\end{enumerate}

Together, these four levels of correlation control (random number synchronization, orbital alignment, deterministic integral decomposition, and consistent classical shadow measurements) establish a robust framework for minimizing stochastic variance in QC-AFQMC force calculations.
This approach enables precise energy differences despite potentially significant absolute energy uncertainties, making accurate force evaluation practical within reasonable computational constraints.

\subsection{Virtual Correlation Energy}\label{sec:vce_theory}

Quantum-classical AFQMC faces a fundamental tension between the dimensionality of quantum resources and the size of realistic chemical systems. Current quantum devices support hundreds of qubits at most, while even medium-sized molecules with 30 to 50 atoms can generate thousands of orbitals in adequate basis sets. The traditional active space approach addresses this mismatch by restricting quantum treatment to a subspace of chemically relevant orbitals, but at the cost of neglecting correlation outside this active space.

Virtual correlation energy offers a resolution to this dilemma by incorporating effects from outside the active space without increasing qubit requirements.\cite{Takeshita2020-dj,Huggins2022-lh,Jiang2025-gc, Zhao2025-fb} The approach begins by partitioning the system into core, active, and virtual spaces, with the trial wavefunction expressed as:

\begin{equation}
    \ket{\Psi_T}=\ket{\Xi^\alpha_c}\otimes\ket{\Xi^\beta_c}\otimes\ket{\Psi_{T,a}}\otimes\ket{0^\alpha_v}\otimes\ket{0^\beta_v}
\end{equation}

Here, $\ket{\Psi_{T,a}}$ represents the quantum-prepared trial state within the active space containing $N_a$ electrons, $\ket{\Xi^{\alpha (\beta)}_c}$ denotes Slater determinants for frozen core $\alpha$ ($\beta$) orbitals with $N^{\alpha (\beta)}_c$ electrons, and $\ket{0^{\alpha (\beta)}_v}$ indicates the vacuum state in the virtual $\alpha$ ($\beta$) orbital space.

This trial wavefunction can be rewritten as a linear combination of Slater determinants within the active space:

\begin{equation}
    \ket{\Psi_T}=\ket{\Xi^\alpha_c}\otimes\ket{\Xi^\beta_c}\otimes\sum_{i}{c_i\ket{\chi_i^\alpha}\otimes\ket{\chi_i^\beta}}\otimes\ket{0^\alpha_v}\otimes\ket{0^\beta_v}
\end{equation}

where $\ket{\chi_i^{\alpha (\beta)}}$ represents the $\alpha$ ($\beta$) component of the $i$-th Slater determinant within the active space.

The overlap between this trial state and a walker defined in the full space can be expressed as:

\begin{equation}
    \sum_i{c_i\braket{\Xi_c^\alpha\Xi_c^\beta\chi^\alpha_i\chi^\beta_i0^\alpha_v0^\beta_v|\phi}}=\sum_i{c_i\mathrm{det}\left(\begin{pmatrix}
\Xi_c^\alpha & 0 & 0 & 0\\
0 & \Xi_c^\beta & 0 & 0 \\
0 & 0 & \chi^\alpha_i & 0 \\
0 & 0 & 0 & \chi^\beta_i \\
0 & 0 & 0 & 0 \\
0 & 0 & 0 & 0 \\
\end{pmatrix}^\dagger\begin{pmatrix}
\phi^\alpha_c & 0 \\
\phi^\alpha_a & 0 \\
\phi^\alpha_v & 0 \\
0 & \phi^\beta_c  \\
0 & \phi^\beta_a  \\
0 & \phi^\beta_v  \\
\end{pmatrix}\right)}
\end{equation}

This simplifies to:

\begin{equation}
    \sum_i{c_i\braket{\Xi_c^\alpha\Xi_c^\beta\chi^\alpha_i\chi^\beta_i|\phi}}=\sum_i{c_i\mathrm{det}\begin{pmatrix}
\Xi_c^{\alpha\dagger}\phi^\alpha_c & 0 \\
0 & \Xi_c^{\beta\dagger}\phi^\beta_c \\
\chi^{\alpha\dagger}_i\phi^\alpha_a & 0 \\
0 & \chi^{\beta\dagger}_i\phi^\beta_a  \\
\end{pmatrix}}
\end{equation}

where $\phi^{\alpha (\beta)}_c$ and $\phi^{\alpha (\beta)}_a$ represent the $N^{\alpha(\beta)}_a+N^{\alpha(\beta)}_c$ column molecular orbital coefficients of the walker, and $\Xi_c^{\alpha(\beta)}$ is diagonal with ones up to the number of $\alpha$($\beta$) core electrons and zeros elsewhere. Crucially, the virtual degrees of freedom no longer appear in this expression.

To further simplify the computation, we perform singular value decomposition on the core orbital components:

\begin{equation}
    \begin{split}
        \Xi_c^{\alpha\dagger}\phi^\alpha_c=U^\alpha_c\Sigma_c^\alpha V^{\alpha\dagger}_c \\
        \Xi_c^{\beta\dagger}\phi^\beta_c=U^\beta_c\Sigma_c^\beta V^{\beta\dagger}_c \\
    \end{split}
\end{equation}

where $U^\alpha_c\in\mathbb{C}^{N^\alpha_c\times N^\alpha_c}$, $V^\alpha_c\in\mathbb{C}^{(N^\alpha_c+N^\alpha_a)\times N^\alpha_c}$, $U^\beta_c\in\mathbb{C}^{N^\beta_c\times N^\beta_c}$, and $V^\beta_c\in\mathbb{C}^{(N^\beta_c+N^\beta_a)\times N^\beta_c}$.

We then define new unitary matrices $U$ and $V$ as:

\begin{equation}
    U=\begin{pmatrix}
U^\alpha_c & 0 & 0 & 0 \\
0 & U^\beta_c & 0 & 0 \\
0 & 0 & I & 0 \\
0 & 0 & 0 & I  \\
\end{pmatrix}
\end{equation}

\begin{equation}
    V=\begin{pmatrix}
V^\alpha_c & 0 & V^{\alpha\prime}_c & 0 \\
0 & V^\beta_c & 0 & V^{\beta\prime}_c \\
\end{pmatrix}
\end{equation}

where $V^{\alpha(\beta)\prime}_c$ are orthonormal columns added to complete the basis.

The overlap can now be rewritten as:

\begin{equation}
    \begin{split}
        \braket{\Xi_c^\alpha\Xi_c^\beta\chi^\alpha_i\chi^\beta_i|\phi}&=\mathrm{det\left(U^\dagger\begin{pmatrix}
\Xi_c^\alpha & 0 & 0 & 0\\
0 & \Xi_c^\beta & 0 & 0 \\
0 & 0 & \chi^\alpha_i & 0 \\
0 & 0 & 0 & \chi^\beta_i \\
\end{pmatrix}^\dagger\begin{pmatrix}
\phi^\alpha_c & 0 \\
0 & \phi^\beta_c  \\
\phi^\alpha_a & 0 \\
0 & \phi^\beta_a  \\
\end{pmatrix}V\right)} / (\mathrm{det}(U^\dagger)\mathrm{det}(V)) \\
&=\mathrm{det}(\Sigma^\alpha_c)\mathrm{det}(\Sigma^\beta_c)\mathrm{det}(\chi^{\alpha\dagger}_i\tilde{\phi}^{\alpha}_a)\mathrm{det}(\chi^{\beta\dagger}_i\tilde{\phi}^{\beta}_a)\mathrm{det}(R^\alpha)\mathrm{det}(R^\beta) / (\mathrm{det}(U^\dagger)\mathrm{det}(V)) \\
    \end{split}
\end{equation}

where $\tilde{\phi}^{\alpha(\beta)}_a$ represents the normalized Slater determinant within the active space, and $\mathrm{det}(R^{\alpha(\beta)})$ is the normalization matrix obtained from QR decomposition of the matrix $\phi^{\alpha(\beta)}_aV^{\alpha(\beta)\prime}$.

The crucial result is that computing the overlap between the trial state and a walker in the full space reduces to evaluating the overlap between the trial wavefunction and a modified determinant in the active space, plus determinant factors from the transformations:

\begin{equation}
    \braket{\Psi_T|\phi}=\mathrm{det}(\Sigma^\alpha_c)\mathrm{det}(\Sigma^\beta_c)\mathrm{det}(R^\alpha)\mathrm{det}(R^\beta)\braket{\Psi_{T,a}|\tilde{\phi}_a}/(\mathrm{det}(U^\dagger)\mathrm{det}(V))
\label{eq:vce_final_overlap}
\end{equation}

This expression enables incorporation of correlation effects from outside the active space without increasing qubit requirements, as the quantum device need only prepare and measure the active space components. The computational overhead remains limited to classical matrix operations on the core and virtual spaces, maintaining the overall polynomial scaling of the method.

The integration of virtual correlation energy with the QC-AFQMC procedure occurs specifically through Eq. \ref{eq:vce_final_overlap}, where the final overlap required for the mixed estimator (Eq. \ref{eq:mixed_estimator}) combines quantum evaluation of the active space overlap $\braket{\Psi_{T,a}|\tilde{\phi}_a}$ with classical computation of the determinant factors. In practice, this approach effectively enables QC-AFQMC to treat significantly larger molecular systems than would be possible with active-space-only methods, while maintaining the crucial quantum advantage for capturing static correlation in the most chemically relevant orbitals. VCE thus resolves the tension between limited quantum resources and the need to capture correlation effects across the full orbital space, enabling accurate treatment of realistic molecular systems within the constraints of current and near-term quantum devices.
\section{Methodology}

The theoretical framework established in Section 2 requires specific methodological implementations to transition from mathematical formalism to practical simulation. This section details three crucial methodological components: (1) active space selection guided by quantum information theory, (2) efficient overlap evaluation using matchgate shadows, and (3) trial state preparation with paired unitary coupled cluster approaches. Together, these elements form the operational core of our QC-AFQMC implementation, balancing quantum resource requirements against computational accuracy. 

\subsection{Active Space Selection}

The selection of appropriate active spaces for QC-AFQMC calculations with virtual correlation energy represents a critical methodological decision, particularly for systems exhibiting significant static correlation.\cite{Stein2016-yg, Stein2019-zw, Toth2020-xm, Goings2022-xb} Rather than relying on arbitrary frontier orbital selection, we implement a systematic approach based on quantum information theory metrics to identify orbitals requiring explicit quantum mechanical treatment.

For this purpose, we primarily make use of tools from the open-source \texttt{ActiveSpaceFinder} library.\cite{HQS-Quantum-SimulationsUnknown-hl}, which is integrated with the electronic structure codes \texttt{PySCF}\cite{Sun2020-mh} and \texttt{Block2}.\cite{Zhai2023-yi} Our procedure begins with preliminary density matrix renormalization group (DMRG) calculations using modest bond dimension (typically $M=250$). From these calculations, we extract the single-orbital entropy for each spatial orbital $p$:

\begin{equation}
S_p = -\sum_{\tau \in \{-, \uparrow, \downarrow, \uparrow\downarrow\}} w_{p,\tau}\ln w_{p,\tau}
\label{eq:single_orbital_entropy}
\end{equation}

This entropy quantifies the quantum entanglement between orbital $p$ and the remainder of the system. The occupation probabilities $w_{p,\tau}$ represent the likelihood of finding orbital $p$ in configuration $\tau$ (empty, $\alpha$-occupied, $\beta$-occupied, or doubly occupied). These probabilities are extracted from the one- and two-particle reduced density matrices (1-RDM $D$ and 2-RDM $d$):

\begin{align}
w_{p,-} &= 1 - D_{pp} - D_{\bar{p}\bar{p}} + d_{pp\bar{p}\bar{p}} \\
w_{p,\uparrow} &= D_{pp} - d_{pp\bar{p}\bar{p}} \\
w_{p,\downarrow} &= D_{\bar{p}\bar{p}} - d_{pp\bar{p}\bar{p}} \\
w_{p,\uparrow\downarrow} &= d_{pp\bar{p}\bar{p}}
\end{align}

where $\bar{p}$ denotes the $\beta$-spin counterpart of the $\alpha$-spin spatial orbital $p$.

While these preliminary DMRG calculations involve modest computational cost, they provide invaluable guidance for active space selection. High entropy values (according to Ref.~\citenum{Stein2016-yg},  $S_p > 0.1 \cdot \ln(4) \approx 0.14$ is typically sufficient) indicate orbitals significantly entangled with the rest of the system --- precisely the orbitals that require inclusion in the active space to capture essential static correlation effects. By examining entropy distributions across multiple geometries relevant to the chemical process under study (e.g., reactant, transition state, product), we identify a consistent active space that captures the essential static correlation across the entire reaction coordinate.

To validate our entropy-based selections, we compare against traditional multireference diagnostics from coupled cluster calculations, including $T_1$, $D_1$,\cite{Lee2009-xz,Leininger2000-jq,Lee2003-th} maximum $\hat{t}_1$- and $\hat{t}_2$-amplitudes,\cite{Cramer2006-bd,Cheng2017-mg,Liakos2011-bo,Feldt2019-vt,Karton2009-av,Radon2014-pu,Gan2006-jz,Margraf2017-av,Cramer1998-jf,West2012-bi,Oliveira2019-xh,Jiang2012-qx,Fogueri2013-nm,Karton2006-mc,Beran2004-tt} and fractional natural orbital occupation numbers (NOONs).\cite{Lee2019-mj,Shee2021-ix,Feldt2019-vt} This cross-validation approach ensures that our active space captures the essential strongly correlated degrees of freedom before proceeding to the more computationally intensive QC-AFQMC calculations.

\subsection{Overlap Evaluation with Matchgate Shadows}

The quantum component of QC-AFQMC centers on measuring overlaps between trial wavefunctions $|\Psi_T\rangle$ and Monte Carlo walkers $|\phi_w\rangle$. This measurement represents the primary quantum advantage in our hybrid quantum-classical approach. Following recent developments,\cite{Wan2023-df,Huang2024-gz, Zhao2025-fb} we implement the matchgate shadow framework, which exploits the algebraic structure of fermionic systems to achieve polynomial-scaling overlap evaluation.

Matchgate circuits consist of nearest-neighbor matchgates---two-qubit operations that preserve fermionic exchange statistics under the Jordan-Wigner mapping. The shadow protocol operates by applying random matchgate circuits $\{U_i\}$ to the trial state $|\Psi_T\rangle$, performing measurements in the computational basis yielding classical bit strings $\{b_i\}$, and reconstructing overlaps through classical post-processing. (See Eq. \ref{eq:shadow_estimator}).

This approach offers significant advantages over direct Hadamard test implementations, requiring fewer circuit executions and providing simultaneous overlap estimates for multiple walkers from the same set of measurements. The number of required shadow samples scales polynomially with system size and inversely with the square of the desired precision, making this approach feasible for systems beyond the reach of exact diagonalization.

Following the efficient implementation framework of Zhao et al.,\cite{Zhao2025-fb}we generate a distribution of random matchgate circuits, apply each to the prepared trial state, measure in the computational basis, and process the results according to Eq.~\ref{eq:shadow_estimator}. For the benchmark calculations presented in this work (H$_6$, N$_2$, CO$_2$), we simulate these measurements classically at the statevector level, corresponding to the infinite shadow limit ($N_s \to \infty$) with zero statistical error in the overlap estimates. This idealized simulation allows us to isolate the effects of AFQMC stochasticity and correlated sampling without introducing additional quantum measurement uncertainties.

\subsection{Trial State Preparation}

The quality of the trial state $|\Psi_T\rangle$ significantly impacts the accuracy and convergence behavior of QC-AFQMC calculations. We implement two primary trial state ansätze: pair unitary coupled cluster with double excitations (upCCD) and its orbital-optimized variant (oo-upCCD). These ansätze balance expressivity with quantum resource requirements, offering sufficient accuracy for strongly correlated systems while maintaining implementability on near-term quantum hardware.

The upCCD ansatz restricts unitary coupled cluster (UCC) to pair excitations, which simultaneously act on spin-up and spin-down electrons in the same spatial orbitals:

\begin{equation}
|\Psi_{\text{upCCD}}\rangle = e^{\hat{T} - \hat{T}^\dagger}|\Phi_0\rangle, \quad \hat{T} = \sum_{i \in \text{occ}} \sum_{a \in \text{virt}} t_{ia} \hat{a}^\dagger_a \hat{a}^\dagger_{\bar{a}} \hat{a}_{\bar{i}} \hat{a}_i
\label{eq:upCCD}
\end{equation}

where $i$ indexes occupied spatial orbitals, $a$ indexes virtual spatial orbitals within the active space, $\hat{a}^\dagger (\hat{a})$ are fermionic creation (annihilation) operators, bars indicate beta spin orbitals, and $|\Phi_0\rangle$ is typically the Hartree-Fock determinant within the active space. This structure significantly reduces the number of parameters and quantum circuit depth compared to full UCCSD, while maintaining the ability to capture essential correlation effects.

The orbital-optimized variant (oo-upCCD) introduces an additional unitary orbital rotation prior to the pair excitation operator:

\begin{equation}
|\Psi_{\text{oo-upCCD}}\rangle = e^{\hat{T} - \hat{T}^\dagger}e^{\hat{\kappa}}|\Phi_0\rangle, \quad \hat{\kappa} = \sum_{p,q \in \text{active}} \kappa_{pq}(\hat{a}^\dagger_p\hat{a}_q - \hat{a}^\dagger_q\hat{a}_p)
\label{eq:oouccd}
\end{equation}

where $\hat{\kappa}$ generates orbital rotations among active orbitals $p, q$, and is anti-Hermitian ($\kappa_{pq} = -\kappa_{qp}^*$). This additional flexibility allows the orbital basis itself to adapt variationally, significantly improving the trial state quality for strongly correlated systems where the initial Hartree-Fock orbitals provide a poor reference. 

Crucially, the orbital rotation $e^{\hat{\kappa}}$ is implemented classically by transforming the one- and two-electron integrals before constructing the quantum circuit for $e^{\hat{T} - \hat{T}^\dagger}$. This hybrid classical-quantum approach enhances trial state quality without increasing quantum circuit depth---a significant practical advantage for near-term quantum implementations.

Both ansätze are optimized using the variational quantum eigensolver (VQE) approach within the selected active space. The resulting optimized trial states then serve as references $|\Phi_T\rangle$ for subsequent QC-AFQMC calculations using the virtual correlation energy framework described in Section 2.2.

This methodological approach---combining information-theoretic active space selection, efficient quantum overlap evaluation, and expressive trial states---forms the foundation for the QC-AFQMC calculations presented in the following sections. The integration of these components enables accurate treatment of strongly correlated systems while maintaining feasibility on near-term quantum hardware.
\section{Results and Discussion}

The quantum computational components of the QC-AFQMC calculations described herein were performed using classical simulators provided by the Qiskit framework.\cite{Javadi-Abhari2024-vi} We emphasize that these results serve primarily to demonstrate the accuracy and effectiveness of the correlated sampling methodology within QC-AFQMC; accordingly, they were not executed on physical quantum hardware, yet. Unless otherwise specified, for the crucial overlap calculations between trial states and walkers (Eq.~\ref{eq:vce_final_overlap}), Qiskit's statevector simulator was used by default. This provides exact wave function evolution, ensuring that the statistical noise observed originates solely from the Monte Carlo sampling and not from the quantum simulation aspect itself. Examples using matchgate shadows use a Qiskit noise-free quantum emulator.

\subsection{Hydrogen Chain Dissociation Studies}
We begin our validation studies with hydrogen chains: prototypical systems exhibiting tunable correlation strength that serve as benchmarks for electronic structure methods. The symmetric stretching of H$_6$ provides a controlled environment to assess QC-AFQMC accuracy across various correlation regimes, from the weakly correlated equilibrium geometry to the strongly correlated dissociation limit.

\begin{figure}[htbp]
\centering
\includegraphics[width=\textwidth]{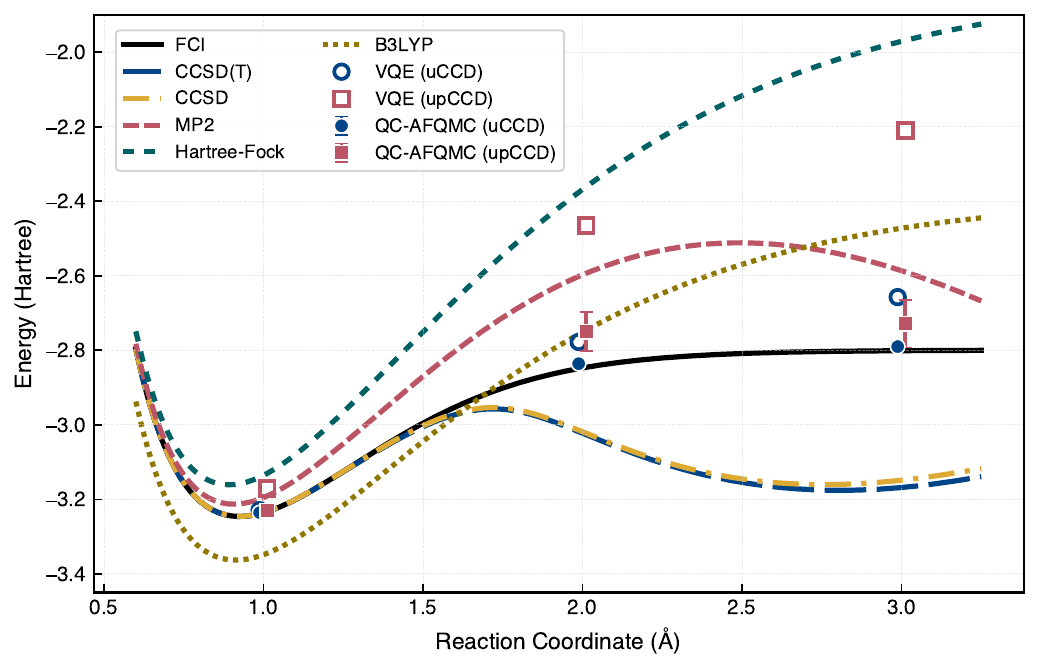}
\caption{Potential energy surface for symmetric H$_6$ dissociation relative to FCI. Comparison of RHF, MP2, CCSD, CCSD(T), DFT (B3LYP), and QC-AFQMC (with upCCD/uCCD trials) results. QC-AFQMC demonstrates high accuracy across the entire dissociation coordinate.}
\label{fig:h6_pes}
\end{figure}

Figure~\ref{fig:h6_pes} presents potential energy surfaces for H$_6$ dissociation computed with various methods relative to Full Configuration Interaction (FCI) results.
Several observations merit emphasis: (1) Hartree-Fock (RHF) theory exhibits the characteristic overestimation of energy in the dissociation limit due to its inability to describe bond breaking correctly. (2) Møller-Plesset perturbation theory (MP2) diverges at stretched geometries due to the breakdown of the perturbative expansion. (3) Coupled Cluster Singles and Doubles (CCSD) remains qualitatively bound but shows increasing error with bond stretching as static correlation grows. (4) CCSD with perturbative triples (CCSD(T)) exhibits non-variational behavior, yielding energies below FCI at intermediate bond lengths---a hallmark of its breakdown in strongly correlated regimes. (5) Standard Density Functional Theory (DFT) functionals (e.g., B3LYP, BP86) qualitatively capture the dissociation curve but often with substantial quantitative errors, particularly in the dissociation limit depending on the functional.

Against this backdrop, QC-AFQMC demonstrates remarkable accuracy across all correlation regimes. Even with the relatively simple upCCD trial wavefunction within an appropriate active space (e.g., 6e$^-$, 6o) combined with virtual correlation energy, QC-AFQMC captures the dissociation curve with errors typically below 5 millihartree compared to FCI. Using the superior uCCD trial wavefunction (or potentially oo-upCCD) further improves accuracy, with errors consistently approaching chemical accuracy (around 1 millihartree) throughout the dissociation coordinate.

This systematic improvement over the underlying trial states illustrates a key advantage of QC-AFQMC: it can leverage modestly accurate trial wavefunctions to produce essentially exact results within the basis set. The initial trial wavefunction need not provide quantitative accuracy itself; its primary role is to guide the projection and mitigate the phase problem (Eq.~\ref{eq:mixed_estimator}). This property makes QC-AFQMC particularly valuable for strongly correlated systems where constructing highly accurate trial states via classical methods or simple VQE ansätze becomes challenging.

\subsection{Nuclear Forces During Bond Stretching}
Computing accurate nuclear forces in strongly correlated regimes provides a more stringent test of QC-AFQMC with correlated sampling. We examine two molecular systems---N$_2$ and linear H$_4$---across various bond lengths, comparing QC-AFQMC force components against FCI, phaseless AFQMC (ph-AFQMC), CCSD, and RHF benchmarks. Forces are computed using the finite difference approach (Eq.~\ref{eq:finite_difference}) with correlated sampling implemented as described in Section 2.3.

Table~\ref{tab:n2_forces} presents force results for N$_2$ at bond lengths ranging from 1.2 \AA~(near equilibrium) to 2.5 \AA~(dissociation region).
Near equilibrium, QC-AFQMC using a (6e$^-$, 6o) active space with virtual correlation energy achieves a force error typically around 0.01 Ha/\AA~compared to FCI, substantially outperforming RHF and demonstrating accuracy comparable to or better than CCSD in this regime. As the bond stretches and correlation strengthens (beyond ~1.8 \AA), QC-AFQMC maintains qualitative correctness and reasonable quantitative accuracy. In contrast, CCSD fails dramatically around 1.8 \AA, predicting a force with the incorrect sign as the single-reference description breaks down.

Phaseless AFQMC (using a single determinant trial) shows a systematic degradation in force accuracy with increasing bond length, with force errors growing substantially in the dissociation limit. This pattern correlates with increasing energy errors observed in ph-AFQMC for stretched N$_2$, reflecting the difficulty of the simple trial wavefunction in constraining the phase problem in this strongly multireference regime. RHF, as expected, fails qualitatively across all stretched geometries. The success of QC-AFQMC here highlights the benefit of using more sophisticated (quantum-prepared) trial wavefunctions combined with the virtual correlation energy approach.

\begin{table}[htbp]
\centering
\caption{Force and energy calculations for N$_2$ / STO-3G at multiple bond lengths\textsuperscript{a}.}
\label{tab:n2_forces}
\begin{threeparttable}
\small
\begin{tabular}{@{}c l S[table-format = -1.4(2)] S[table-format = -1.6(2)]  S[table-format = -3.4(2)] S[table-format = -1.6(2)]@{}}
\toprule
& & \multicolumn{2}{c}{\textbf{Force} \si{\hartree\per\angstrom}}  & \multicolumn{2}{c}{\textbf{Energy} \si{\hartree}} \\
\cmidrule(lr){3-4} \cmidrule(lr){5-6}
\textbf{Bond Length} & \textbf{Method} & \multicolumn{1}{c}{\textbf{Value}} & \multicolumn{1}{c}{\textbf{$\Delta$F$_{\text{FCI}}$}} & \multicolumn{1}{c}{\textbf{Value}} & \multicolumn{1}{c}{\textbf{$\Delta$E$_{\text{FCI}}$}} \\
\addlinespace[0.5ex]
\multirow{5}{*}{{1.2 \si{\angstrom}}} & FCI\textsuperscript{\textit{ref}} & -0.0235 & {} & -107.6773 & {} \\
& QC-AFQMC & -0.0338(21) & 0.010261 & -107.6724(9) & -0.004932 \\
& ph-AFQMC & -0.1010(60) & 0.077509 & -107.6633(25) & -0.014058 \\
& CCSD & -0.0467 & 0.023214 & -107.6714 & -0.005901 \\
& RHF & -0.3631 & 0.339578 & -107.4878 & -0.189556 \\
\addlinespace[1ex]
\multirow{5}{*}{{1.6 \si{\angstrom}}} & FCI\textsuperscript{\textit{ref}} & -0.3695 & {} & -107.5421 & {} \\
& QC-AFQMC & -0.4160(13) & 0.046099 & -107.5351(10) & -0.006941 \\
& ph-AFQMC & -0.6678(70) & 0.298280 & -107.4725(48) & -0.069592 \\
& CCSD & -0.3291 & -0.040457 & -107.5294 & -0.012738 \\
& RHF & -0.8722 & 0.502633 & -107.1848 & -0.357243 \\
\addlinespace[1ex]
\multirow{5}{*}{{2.0 \si{\angstrom}}} & FCI\textsuperscript{\textit{ref}} & -0.0842 & {} & -107.4552 & {} \\
& QC-AFQMC & -0.1786(69) & 0.094331 & -107.4274(13) & -0.027754 \\
& ph-AFQMC & -0.5563(58) & 0.472020 & -107.2214(53) & -0.233794 \\
& CCSD & 0.3080 & -0.392268 & -107.5570 & 0.101829 \\
& RHF & -0.6642 & 0.579934 & -106.8715 & -0.583652 \\
\addlinespace[1ex]
\multirow{5}{*}{{2.5 \si{\angstrom}}} & FCI\textsuperscript{\textit{ref}} & -0.0074 & {} & -107.4404 & {} \\
& QC-AFQMC & -0.0182(38) & 0.010841 & -107.3950(24) & -0.045416 \\
& ph-AFQMC & -0.3301(39) & 0.322707 & -107.0155(65) & -0.424955 \\
& CCSD & -0.1013 & 0.093893 & -107.1815 & -0.258885 \\
& RHF & -0.3725 & 0.365099 & -106.6170 & -0.823457 \\
\bottomrule
\end{tabular}
\begin{tablenotes}
\footnotesize
\item[a] Both AFQMC calculations employed 1024 walkers over 150 blocks with 10 steps per block ($\Delta t = 0.01$, $\Delta x = 10^{-6}$). Forces represent gradient for one N atom along bond axis. QC-AFQMC uses a (6e$^-$, 6o) active space with upCCD trial and virtual correlation energy, using statevector simulator for the overlaps. All other methods utilize the full space. Uncertainties shown in parentheses correspond to the last digits.
\end{tablenotes}
\end{threeparttable}
\end{table}
The linear H$_4$ chain, with symmetric H-H stretching, presents increasing electronic structure challenges as the bond length grows. Table~\ref{tab:h4_forces} shows force results for H$_4$. At equilibrium (H-H $\approx$ 1.0\AA), both QC-AFQMC and ph-AFQMC achieve close agreement with FCI and CCSD. However, as the separation increases to $R = 2.0$\AA, the system becomes more strongly correlated. The force errors for QC-AFQMC grow and ph-AFQMC exhibits even larger deviations from FCI. Despite the growing errors, QC-AFQMC still demonstrates a significant advantage over CCSD and ph-AFQMC. These latter methods struggle with the multireference character at long bond distances, underscoring the inherent limitations of simpler trial wave functions in strongly correlated regimes.

\begin{table}[htbp]
\centering
\caption{Force and energy calculations for H$_4$ linear chain / STO-3G at multiple bond lengths\textsuperscript{a}.}
\label{tab:h4_forces}
\begin{threeparttable}
\small
\begin{tabular}{@{}c l S[table-format = -1.3(2)] S[table-format = -1.3] S[table-format = -1.3(2)] S[table-format = -1.3]@{}}
\toprule
& & \multicolumn{2}{c}{\textbf{Force} \si{\hartree\per\angstrom}}  & \multicolumn{2}{c}{\textbf{Energy} \si{\hartree}} \\
\cmidrule(lr){3-4} \cmidrule(lr){5-6}
\textbf{Bond Length} & \textbf{Method} & \multicolumn{1}{c}{\textbf{Value}} & \multicolumn{1}{c}{\textbf{$\Delta$F$_{\text{FCI}}$}} & \multicolumn{1}{c}{\textbf{Value}} & \multicolumn{1}{c}{\textbf{$\Delta$E$_{\text{FCI}}$}} \\
\addlinespace[0.5ex]
\multirow{6}{*}{{1.0 \si{\angstrom}}} & FCI\textsuperscript{\textit{ref}} & 0.169 & 0.000 & -2.166 & 0.000 \\
& QC-AFQMC (statevector) & 0.171(2) & -0.003 & -2.164(2) & -0.003 \\
& QC-AFQMC (matchgate) & 0.175(3) & -0.006 & -2.167(4) & 0.000 \\
& ph-AFQMC & 0.188(3) & -0.020 & -2.158(4) & -0.009 \\
& CCSD & 0.169 & 0.000 & -2.166 & 0.000 \\
& RHF & 0.237 & -0.068 & -2.099 & -0.068 \\
\addlinespace[1ex]
\multirow{6}{*}{{1.5 \si{\angstrom}}} & FCI\textsuperscript{\textit{ref}} & 0.144 & 0.000 & -1.996 & 0.000 \\
& QC-AFQMC (statevector) & 0.177(7) & -0.033 & -1.976(7) & -0.020 \\
& QC-AFQMC (matchgate) & 0.162(10) & -0.018 & -1.978(7) & -0.018 \\
& ph-AFQMC & 0.209(6) & -0.064 & -1.949(10) & -0.047 \\
& CCSD & 0.132 & 0.012 & -1.998 & 0.001 \\
& RHF & 0.277 & -0.133 & -1.829 & -0.167 \\
\addlinespace[1ex]
\multirow{6}{*}{{2.0 \si{\angstrom}}} & FCI\textsuperscript{\textit{ref}} & 0.045 & 0.000 & -1.898 & 0.000 \\
& QC-AFQMC (statevector) & 0.033(29) & 0.011 & -1.864(17) & -0.034 \\
& QC-AFQMC (matchgate) & 0.068(11) & -0.024 & -1.851(15) & -0.046 \\
& ph-AFQMC & 0.136(5) & -0.091 & -1.749(12) & -0.149 \\
& CCSD & -0.009 & 0.053 & -1.916 & 0.018 \\
& RHF & 0.195 & -0.151 & -1.576 & -0.322 \\
\bottomrule
\end{tabular}
\begin{tablenotes}
\footnotesize
\item[a] Both AFQMC calculations employed 256 walkers over 80 blocks with 10 steps per block ($\Delta t = 0.02$, $\Delta x = 10^{-5}$). Forces represent gradient for the terminal H atom along the chain axis. QC-AFQMC uses upCCD trial state with no active space reduction; overlaps computed either exactly (statevector) or using matchgate shadows (21,080 shadows). All methods utilize the full space. Uncertainties shown in parentheses correspond to the last digits of the main number.
\end{tablenotes}
\end{threeparttable}
\end{table}

An important observation from Tables~\ref{tab:n2_forces} and~\ref{tab:h4_forces} is that even when using correlated sampling, the statistical error in the force generally increases with bond length. This trend reflects the growing challenge of maintaining perfect correlation as electronic structure changes become more dramatic with increasing geometric displacement. Nevertheless, for both systems, the correlation efficiency remains sufficient to enable meaningful force evaluation across the entire dissociation coordinate, validating the robustness of our correlated sampling implementation for strongly correlated systems.

\subsection{The Effect of Orbital Optimization on Trial States}
\begin{figure}[htbp]
\centering
\includegraphics[width=0.8\textwidth]{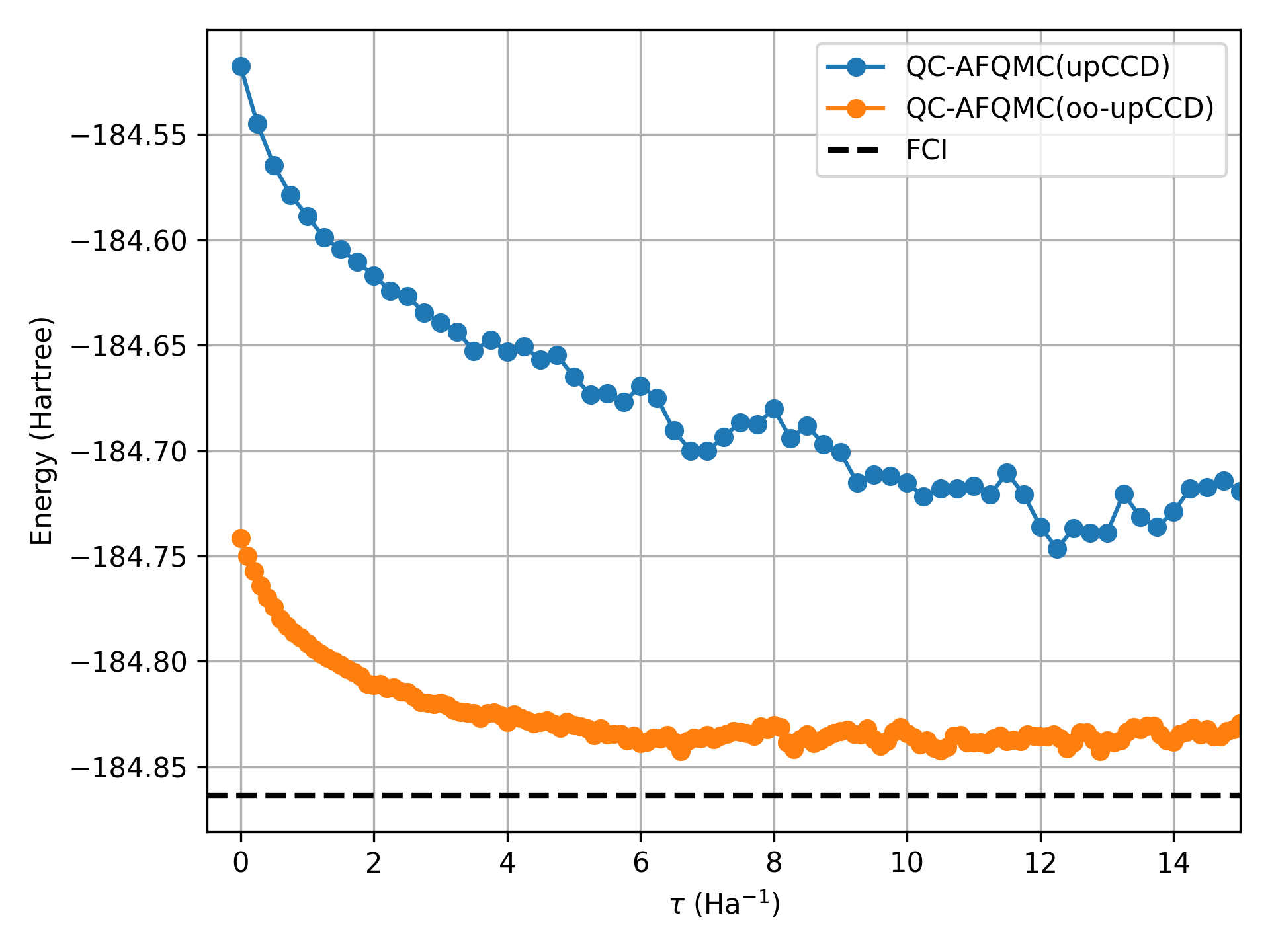}

\caption{Comparison of QC-AFQMC energy convergence for stretched CO$_2$ ($R_{\text{C-O}} = 2.0$ \AA) using standard upCCD vs. orbital-optimized (oo-upCCD) trial wavefunctions. The oo-upCCD trial leads to lower absolute energy and more stable convergence. The black dashed line indicates the FCI reference energy.}
\label{fig:co2_trial_compare}
\end{figure}

The influence of trial wavefunction quality becomes particularly apparent in the CO$_2$ system at stretched geometries. Figure~\ref{fig:co2_trial_compare} compares QC-AFQMC energy convergence using standard upCCD\cite{Elfving2021-pi,Zhao2023-ck,O-Brien2023-ft} versus orbital-optimized upCCD (oo-upCCD)\cite{Zhao2023-ck} trial wavefunctions for CO$_2$ at a stretched C-O bond length of 2.0 \AA.

The oo-upCCD trial wavefunction itself exhibits significantly lower variational energy (approximately 200 millihartrees below upCCD), indicating its superior description of the electronic structure in this strongly correlated regime where the RHF orbitals are a poor starting point.

This trial wavefunction improvement propagates through the QC-AFQMC calculation. The projection using the oo-upCCD trial yields final energies much closer to the FCI reference compared to using the standard upCCD trial. The convergence profiles (energy vs. imaginary time) reveal two significant effects: (1) oo-upCCD provides a better starting point (lower mixed estimator energy at $\tau=0$), and (2) the projection path guided by oo-upCCD encounters less severe phase problems, resulting in smaller statistical fluctuations and more stable convergence towards the ground state.

\begin{figure}[htbp]
\centering
 \includegraphics[width=0.8\textwidth]{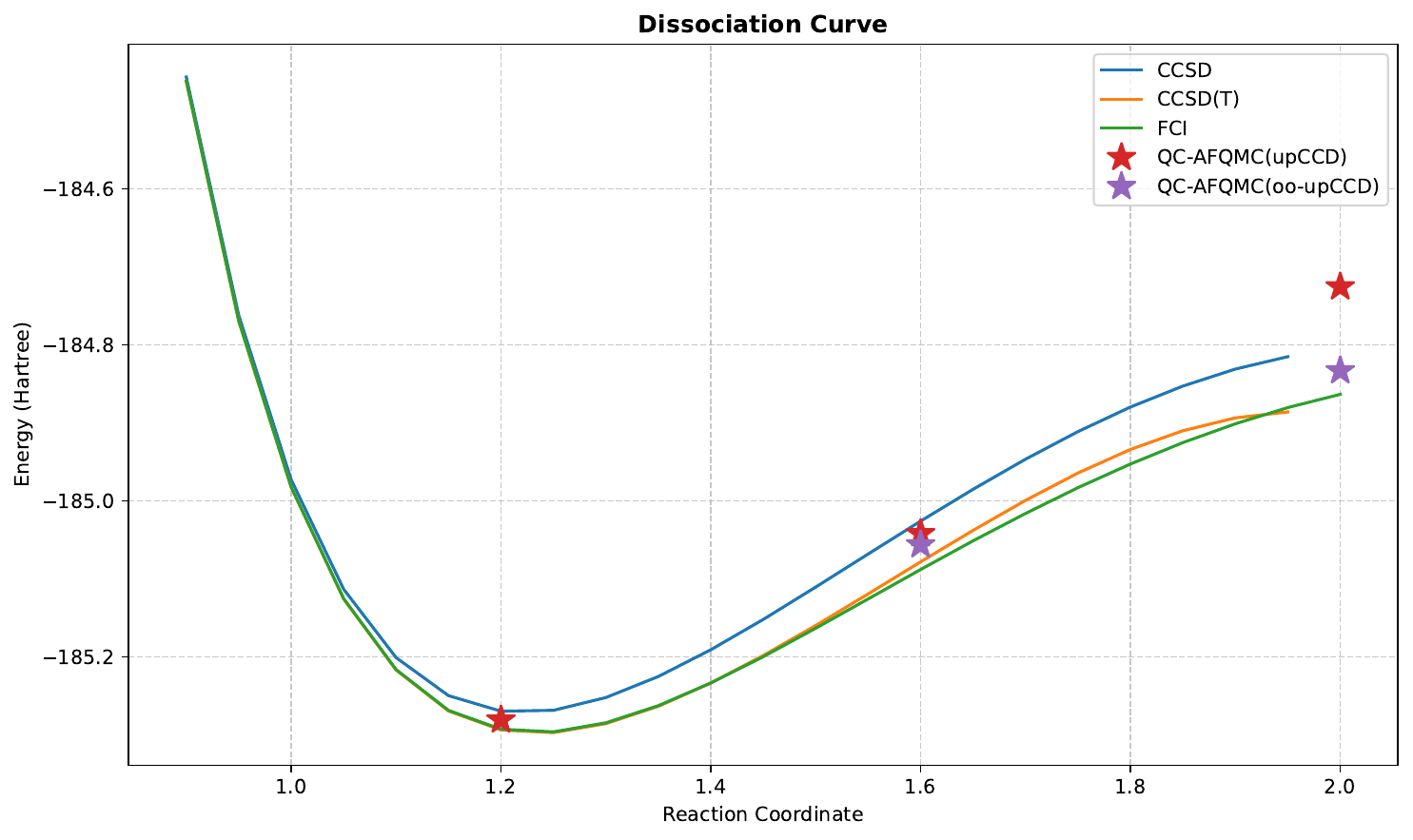}

\caption{Potential energy surface for symmetric CO$_2$ dissociation. Comparison showing the failure of CCSD/CCSD(T) at large distances and the improved accuracy of QC-AFQMC using an oo-upCCD trial wavefunction compared to a standard upCCD trial, relative to FCI reference energies. Simulations done using the Qiskit Aer simulator~\cite{Javadi-Abhari2024-vi}.}
\label{fig:co2_pes_ootrial}
\end{figure}

Figure~\ref{fig:co2_pes_ootrial} extends this comparison across the entire CO$_2$ dissociation coordinate, plotting potential energy surfaces from various methods, including QC-AFQMC with both upCCD and oo-upCCD trial wavefunctions.

Notably, standard CCSD and CCSD(T) methods fail to converge beyond a certain bond distance (approximately 1.95 \AA~for CO$_2$) due to the strongly multireference character of the stretched molecule. In contrast, QC-AFQMC, especially when using the enhanced oo-upCCD trial wavefunction combined with virtual correlation energy, maintains accuracy throughout the coordinate, providing reliable energetics even in regimes where conventional coupled cluster methods break down completely.

The orbital optimization incorporates important static correlation effects directly into the reference orbitals used by the subsequent pair excitation operator (Eq.~\ref{eq:oouccd}). This effectively ``pre-conditions'' the problem, making it easier for the relatively simple upCCD operator to capture the remaining correlation. This approach does not increase quantum resource requirements for the overlap estimation step, as the orbital rotation is implemented classically by transforming the integrals before constructing the quantum circuit for the UCC part. The dramatic improvement observed underscores the importance of leveraging classical preprocessing and algorithm co-design to enhance the performance of quantum algorithms in computational chemistry.

\subsection[Application to MEA-CO2 Reaction]{Application to MEA-CO$_2$ Reaction}
Having validated our methodology on benchmark systems, we turn to a chemically relevant application: the reaction between monoethanolamine (MEA) and CO$_2$. This reaction is a prototype for amine-based carbon capture processes. It proceeds through a zwitterionic mechanism, involving nucleophilic attack of the MEA nitrogen on the CO$_2$ carbon, followed by proton transfer to form a carbamate species. Accurate energetics, particularly barrier heights, are crucial for understanding and optimizing capture efficiency.\cite{da-Silva2004-fv, Xie2010-wu, Rochelle2012-bt, Lin2024-qs}

\begin{figure}[htbp]
\centering
 \includegraphics[width=0.8\textwidth]{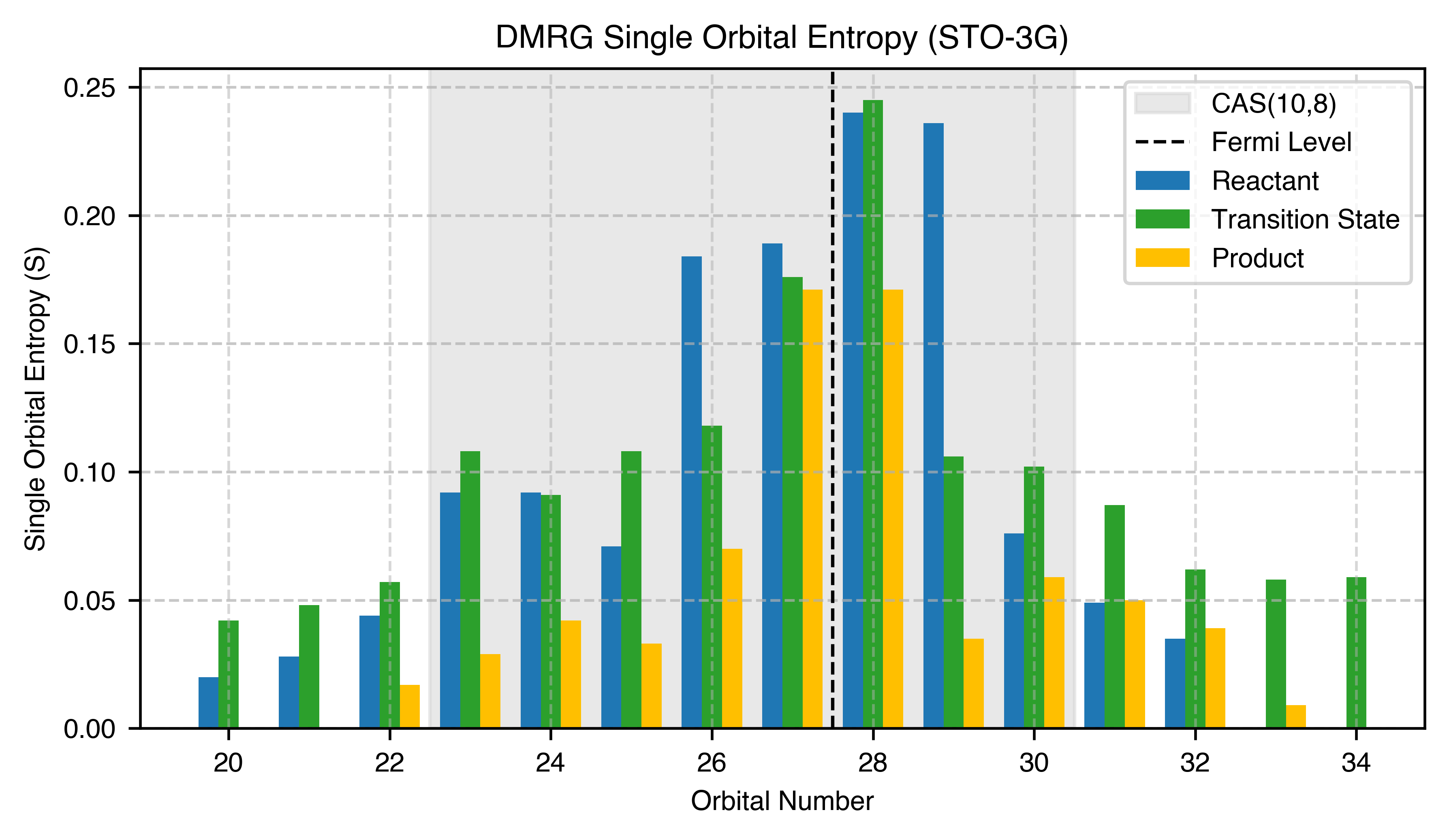}

\caption{Single-orbital entropies ($S_p$) calculated via DMRG for the MEA-CO$_2$ system at different points along the reaction coordinate (reactant, transition state, product). Orbitals with high entropy ($S_p > 0.1$) are candidates for the active space. This guides the selection of a (10e$^-$, 8o) active space for QC-AFQMC calculations.}
\label{fig:mea_entropy}
\end{figure}

Active space selection for this larger, more complex system follows our quantum information theory approach described in the Methodology section. Figure~\ref{fig:mea_entropy} presents single-orbital entropies calculated from preliminary DMRG calculations across key points on the reaction coordinate (reactant complex, transition state, product complex).

Orbitals near the Fermi level, particularly those involved in the N-C bond formation and charge rearrangement (e.g., N lone pair, CO$_2$ $\pi^*$ orbitals), consistently exhibit the highest entanglement entropy, with several showing values exceeding the threshold of 0.14. Based on this analysis, validated through multireference diagnostics (see Table~\ref{tab:mea_diagnostics}), we select a (10e$^-$, 8o) active space, capturing the dominant static correlation effects while remaining tractable for quantum simulation via QC-AFQMC with matchgate shadows.

\begin{table}[htbp]
\centering
\caption{Multireference diagnostics for MEA-CO$_2$ reaction intermediates across different candidate active spaces, used to validate active space selection. The (10e$^-$, 8o) space captures the essential static correlation while remaining computationally tractable.}
\label{tab:mea_diagnostics}
\begin{tabular}{lccccc}
\toprule
System & T$_1$ & D$_1$ & $\max(|t_1|)$ & $\max(|t_2|)$ & \# frac NOONs \\
\midrule
\multicolumn{6}{l}{\textbf{(4e$^-$,4o) Active Space}} \\
Reactant & 0.0001 & 0.0001 & 0.0002 & 0.0027 & 0 \\
Transition State & 0.0048 & 0.0066 & 0.0177 & 0.1700 & 2 \\
Product & 0.0015 & 0.0021 & 0.0055 & 0.1170 & 2 \\
\midrule
\multicolumn{6}{l}{\textbf{(10e$^-$,8o) Active Space}} \\
Reactant & 0.0007 & 0.0012 & 0.0098 & 0.1152 & 4 \\
Transition State & 0.0018 & 0.0042 & 0.0186 & 0.1647 & 2 \\
Product & 0.0009 & 0.0019 & 0.0101 & 0.1232 & 2 \\
\midrule
\multicolumn{6}{l}{\textbf{(16e$^-$,14o) Active Space}} \\
Reactant & 0.0003 & 0.0010 & 0.0095 & 0.1142 & 4 \\
Transition State & 0.0011 & 0.0039 & 0.0369 & 0.1623 & 2 \\
Product & 0.0004 & 0.0017 & 0.0113 & 0.1201 & 2 \\
\bottomrule
\end{tabular}
\end{table}

\begin{figure}[htbp]
\centering
 \includegraphics[width=0.8\textwidth]{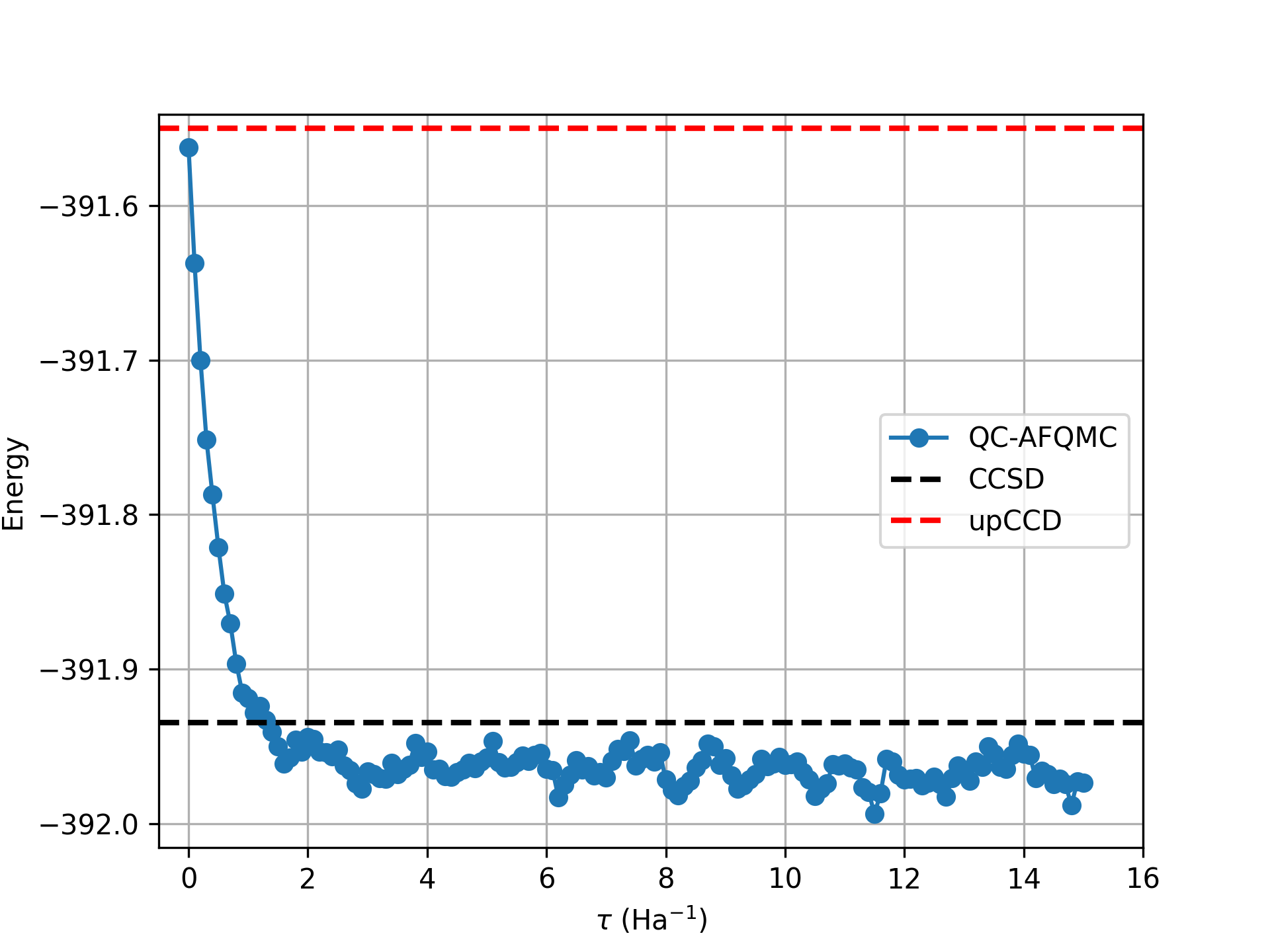}

\caption{QC-AFQMC energy convergence (mixed estimator vs. imaginary time $\tau$) for the MEA-CO$_2$ reactant state. Comparison with the active-space VQE trial energy (upCCD) and full-space CCSD energy shows the energy lowering due to projection and inclusion of virtual correlation. Simulations done using the Qiskit Aer simulator~\cite{Javadi-Abhari2024-vi}.}
\label{fig:mea_convergence}
\end{figure}

Figure~\ref{fig:mea_convergence} tracks the QC-AFQMC energy convergence for the MEA-CO$_2$ reactant state, comparing the final projected energy with the VQE energy of the trial wavefunction (upCCD) within the active space and with a full-space classical CCSD calculation.

The QC-AFQMC energy, incorporating dynamic correlation from the virtual space through the virtual correlation energy approach (Eq. \ref{eq:vce_final_overlap}), falls significantly below the active-space-only VQE energy and agrees well with the CCSD reference. This confirms that the chosen active space and trial wavefunction adequately capture static correlation, and the virtual correlation treatment successfully incorporates the remaining dynamic correlation from outside the active space. The convergence behavior also demonstrates the stability of the QC-AFQMC projection, with the mixed estimator energy (Eq. \ref{eq:mixed_estimator}) smoothly approaching its asymptotic value as imaginary time increases.

\begin{figure}[htbp]
\centering
\includegraphics[width=0.8\textwidth]{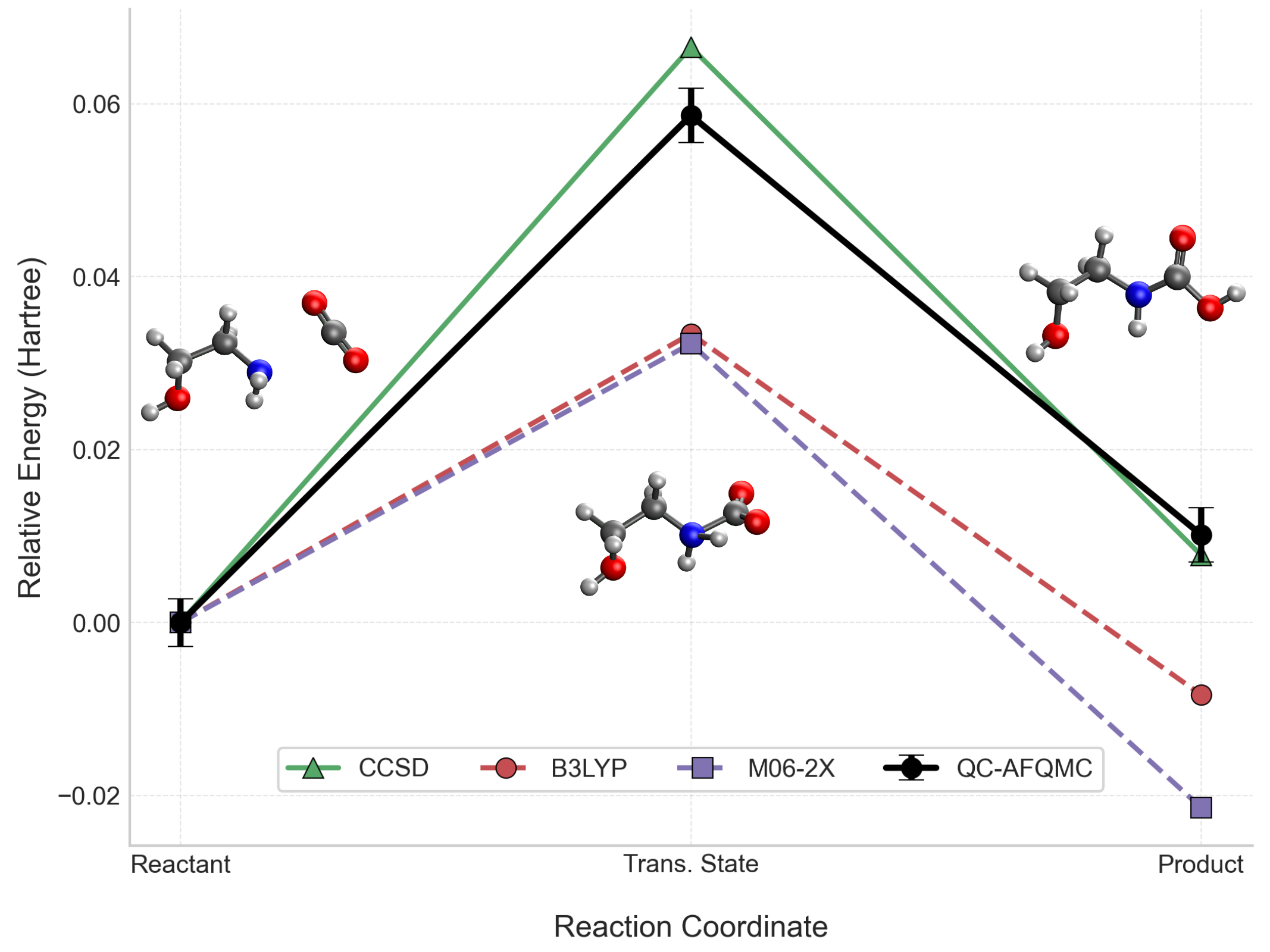}
\caption{Reaction energy profile for the MEA-CO$_2$ reaction (Reactant $\to$ TS $\to$ Product). Comparison of QC-AFQMC results with DFT (B3LYP, M06-2X) and CCSD. The results demonstrate that some DFT functionals significantly underestimate the activation barrier and reaction energy, while QC-AFQMC agrees well with CCSD.}
\label{fig:mea_reaction_profile}
\end{figure}

Figure~\ref{fig:mea_reaction_profile} presents the computed reaction energy profile, including the activation barrier, using QC-AFQMC compared to DFT methods (B3LYP, M06-2X) and CCSD.

DFT functionals show significant variation in the predicted barrier height, with some functionals (e.g., B3LYP) underestimating the barrier by more than 50\% compared to the QC-AFQMC and CCSD results. This highlights the challenges DFT methods face when dealing with charge transfer processes and the importance of accurate electron correlation treatment. The close agreement between QC-AFQMC and CCSD for this reaction suggests that while the system involves significant charge rearrangement, it does not exhibit extreme multireference character at the transition state---consistent with the moderate $\mathrm{T}_1$ diagnostics shown in Table~\ref{tab:mea_diagnostics}. This validates our active space selection and confirms the effectiveness of the QC-AFQMC approach with correlated sampling for studying chemical reactions of industrial relevance.

In addition to the energetic profile, we examined the nuclear forces at the MEA-CO$_2$ transition state to further assess the performance of QC-AFQMC. Specifically, we calculated the force on the carbon atom involved in the forming C-N bond, along the C-N stretch direction. As before, the calculations employed an STO-3G basis set and a (10e$^-$, 8o) active space for the AFQMC methods, with overlaps determined using the exact statevector simulator. As FCI results are intractable for this system, CCSD, which shows high consistency with QC-AFQMC for energies (see, e.g., Fig.~\ref{fig:mea_reaction_profile}), serves as the reference for comparing forces and energy errors. Table~\ref{tab:mea_co2_forces_detail} summarizes these force calculations.

\begin{table}[htbp]
\centering
\caption{Force and energy calculations for MEA-CO$_2$ transition state using CAS(10e$^{-}$,8o)/STO-3G reference geometry\textsuperscript{a}.}
\label{tab:mea_co2_forces_detail}
\begin{threeparttable}
\small
\begin{tabular}{@{}l S[table-format = -1.4(2)] S[table-format=-1.6] @{\hspace{1.5em}} S[table-format = -3.2(1)] S[table-format=-1.6]@{}}
\toprule
& \multicolumn{2}{c}{\textbf{Force}} & \multicolumn{2}{c}{\textbf{Energy}} \\
\cmidrule(lr){2-3} \cmidrule(lr){4-5}
\textbf{Method} & {\textbf{Value} \si{\hartree\per\angstrom}} & {\textbf{$\Delta$F$_{\text{CCSD}}$} \si{\hartree\per\angstrom}} & {\textbf{Value} \si{\hartree}} & {\textbf{$\Delta$E$_{\text{CCSD}}$} \si{\hartree}} \\
\midrule
CCSD\textsuperscript{\textit{ref}} & -0.022 & {} & -391.868 & {} \\
\addlinespace[0.5ex]
QC-AFQMC & -0.020(4) & -0.002 & -391.870(24) & 0.002 \\
ph-AFQMC & -0.015(7) & -0.006 & -391.855(26) & -0.013 \\
RHF & 0.005 & -0.026 & -391.435 & -0.433 \\
\bottomrule
\end{tabular}
\begin{tablenotes}
\footnotesize
\item[a] Both AFQMC calculations employed 512 walkers over 50 blocks with 10 steps per block ($\Delta t = 0.02$, $\Delta x = 10^{-5}$). Forces represent C atom displacement along C--N bond axis.QC-AFQMC uses a (10e$^-$, 8o) active space with upCCD trial and virtual correlation energy, using the statevector simulator for the overlaps. Uncertainties shown in parentheses correspond to the last digits.
\end{tablenotes}
\end{threeparttable}
\end{table}

The results presented in Table~\ref{tab:mea_co2_forces_detail} indicate that CCSD predicts a force of $\approx -0.022$ Ha/\AA~on the carbon atom. The QC-AFQMC method, with its (10e$^-$, 8o) active space, yields a force of $\approx -0.020$ Ha/\AA, demonstrating excellent agreement with CCSD. The absolute deviation from the CCSD force ($|\Delta F_{\text{CCSD}}|$) is around $\approx 0.002$ Ha/\AA, and its energy at the reference geometry is 2 mHa lower than CCSD, albeit with a statistical uncertainty of $\pm$24 mHa. In comparison, the classical ph-AFQMC method, using a single Slater determinant trial, results in a force of $\approx -0.015$ Ha/\AA. This value has a larger deviation from CCSD ($|\Delta F_{\text{CCSD}}| = 0.006$ Ha/\AA) and an energy 13 mHa higher than CCSD (though note the statistical uncertainty $\pm$26 mHa). The RHF method, as anticipated, shows the largest discrepancy, predicting a force of $\approx +0.005$ Ha/\AA, which is the wrong sign entirely. These force calculations further corroborate the utility and accuracy of the QC-AFQMC approach and highlight its advantages over classical ph-AFQMC when a more sophisticated trial wavefunction and active space treatment are employed.

This application demonstrates the practical utility of QC-AFQMC with correlated sampling for studying realistic chemical reactions. By focusing quantum resources on the orbitals most critical for describing bond making/breaking and static correlation, while efficiently incorporating dynamic correlation from the larger virtual space, the method provides accurate energetics across reaction coordinates, even for systems challenging for traditional approaches like DFT. The correlated sampling implementation enables precise force evaluation without prohibitive computational cost, making it feasible to explore potential energy surfaces and reaction mechanisms for industrially relevant processes like carbon capture.
\section{Conclusion}
We have developed and implemented a comprehensive correlated sampling approach for computing nuclear forces within the QC-AFQMC framework. By systematically controlling stochastic error through synchronized random number streams, a rigorous orbital alignment protocol across geometries, the enforcement of deterministic two-electron integral decompositions, and the consistent application of classical shadow measurement ensembles derived from the reference geometry, we achieve precise force evaluation despite the inherent statistical nature of the underlying Monte Carlo process. The key is maximizing the statistical correlation ($\rho \to 1$ in Eq.~\ref{eq:force_variance_correlated}) between energy evaluations at finitely displaced geometries, which dramatically reduces the variance in the computed force components.

Our validation studies included an examination of energies for H$_6$, detailed force calculations for N$_2$ and linear H$_4$, and an investigation into the impact of advanced trial wavefunctions on the accurate energetic description of CO$_2$. From this we demonstrate that QC-AFQMC with correlated sampling provides high accuracy across varying correlation regimes, from weakly correlated equilibrium geometries to strongly correlated stretched bonds where traditional single-reference methods like CCSD can fail qualitatively. The systematic improvement over both Hartree-Fock and standard coupled cluster methods highlights QC-AFQMC's potential for a balanced treatment of static and dynamic correlation, enabled by active space selection and virtual correlation energy.

The CO$_2$ system revealed the crucial role of trial wavefunction quality in handling challenging electronic structures involving multiple bond dissociations. While standard upCCD trial wavefunctions proved adequate for single bond breaking (N$_2$), orbital-optimized variants (oo-upCCD) dramatically improved accuracy for stretched CO$_2$ without increasing quantum resource requirements for the overlap estimation step. This finding emphasizes the importance of classical preprocessing and adaptive ans{\"a}tze in quantum algorithm design and suggests promising directions for developing more robust and efficient trial wavefunctions.
Our application to the MEA-CO$_2$ reaction demonstrates the practical utility of these methods for chemically relevant systems. The successful combination of quantum information theory-based active space selection (using DMRG-derived entropies), the virtual correlation energy technique for recovering dynamic correlation outside the active space, and efficient overlap evaluation (via matchgate shadows) establishes a powerful framework for accurate quantum chemical calculations on near-term quantum resources. The ability to compute reliable forces opens the door to use QC-AFQMC for several applications. This includes geometry optimizations, reaction path following, and metadynamics simulations to explore configuration spaces based on machine-learned force-fields.\cite{Guan2023-ii}
Several methodological extensions merit further investigation. First, exploring analytical implementations of QC-AFQMC forces, potentially based on extensions of the Hellmann-Feynman theorem or Lagrangian techniques adapted to the stochastic and phaseless nature of the method, could offer computational advantages over finite difference approaches. Second, developing more sophisticated trial state ans{\"a}tze that can be efficiently prepared on quantum hardware and effectively capture complex correlation patterns could further improve accuracy and robustness. Finally, integrating QC-AFQMC forces with standard geometry optimization and reaction path algorithms would enable comprehensive exploration of potential energy surfaces with quantum-level accuracy for strongly correlated systems.
In conclusion, correlated sampling significantly enhances the capabilities of QC-AFQMC by enabling precise and statistically controlled force evaluation. This development extends the applicability of this promising quantum-classical algorithm beyond simple energy calculations to force-dependent properties, reaction mechanisms, and potentially molecular dynamics simulations, establishing quantum advantage for accurately treating strongly correlated chemical systems that challenge conventional electronic structure methods.

\begin{acknowledgement}
We thank the Hyundai Motor Company for funding this research through the Hyundai-IonQ Joint Quantum Computing Research Project.
\end{acknowledgement}

\section{Supporting Information}

\subsection{Schematics and Technical Implementation Details}

The following schematics provide visual representations of the key algorithmic components described in this work.

\subsubsection{Correlated Sampling Framework}
\begin{figure}[htb]
    \centering
    \resizebox{0.375\textwidth}{!}{
        \begin{tikzpicture}[
    font=\sffamily\Large,
    node distance=1.35cm and 1.8cm,
    box/.style={
        rectangle, 
        draw,
        text width=8cm,
        minimum height=1.2cm,
        align=center,
        inner sep=5pt
    },
    widebox/.style={
        rectangle, 
        draw,
        text width=18cm,
        minimum height=1.2cm,
        align=left,
        inner sep=10pt
    },
    note/.style={
        rectangle,
        draw,
        dashed,
        text width=6.5cm,
        align=left,
        inner sep=5pt
    },
    arrow/.style={
        -{Stealth[length=4mm, width=2mm]},
        thick
    }
]

\node[box] (start) {
    \textbf{Input:} Molecular configuration\\
    geometry, basis set, active space
};

\node[box] (ref_opt) [below=0.7cm of start] {
    \textbf{Reference State Generation:}\\[0.2cm]
    VQE optimization of $|\Phi_T\rangle$\\
    Classical shadow generation
};

\node[box] (store_ref) [below=0.7cm of ref_opt] {
    \textbf{Structural Reference:}\\[0.2cm]
    Cholesky pivots $(p_k,q_k)$\\
    Orbital alignment protocol
};

\node[box, text width=11cm, align=center] (combined_box) [below=0.7cm of store_ref] {
    \textbf{Synchronized Overlap Calculations:}\\[0.3cm]
    $\mathbf{R} \pm \delta\mathbf{e}_i$ with preserved:\\[0.1cm]
    • shadows \quad • seeds \quad • pivots \quad • orbitals\\[0.3cm]
    Quantum synchronization enforced via:\\[0.1cm]
    • single shadow set \\
    • fixed circuit parameters\\
    • consistent tomography
};

\node[box] (collect) [below=0.7cm of combined_box] {
    \textbf{Deterministic Energy Evaluation:}\\[0.2cm]
    $E = \frac{\langle\Phi_T|\hat{H}|\Phi\rangle}{\langle\Phi_T|\Phi\rangle}$\\
    via synchronized parallel sampling
};

\node[box] (force) [below=0.7cm of collect] {
    \textbf{Force Computation:}\\[0.2cm]
    $F_i = -\frac{E(\mathbf{R} + \delta\mathbf{e}_i) - E(\mathbf{R} - \delta\mathbf{e}_i)}{2\delta}$
};

\draw[arrow] (start.south) -- (ref_opt.north);
\draw[arrow] (ref_opt.south) -- (store_ref.north);
\draw[arrow] (store_ref.south) -- (combined_box.north);
\draw[arrow] (combined_box.south) -- (collect.north);
\draw[arrow] (collect.south) -- (force.north);

\end{tikzpicture}
    }
    \caption{Schematic representation of the correlated sampling framework for nuclear gradient calculations. The workflow maintains consistent random seeds, Cholesky pivots, and orbital alignments across displaced geometries to ensure maximal correlation between QMC runs.}
    \label{fig:correlated-sampling}
\end{figure}
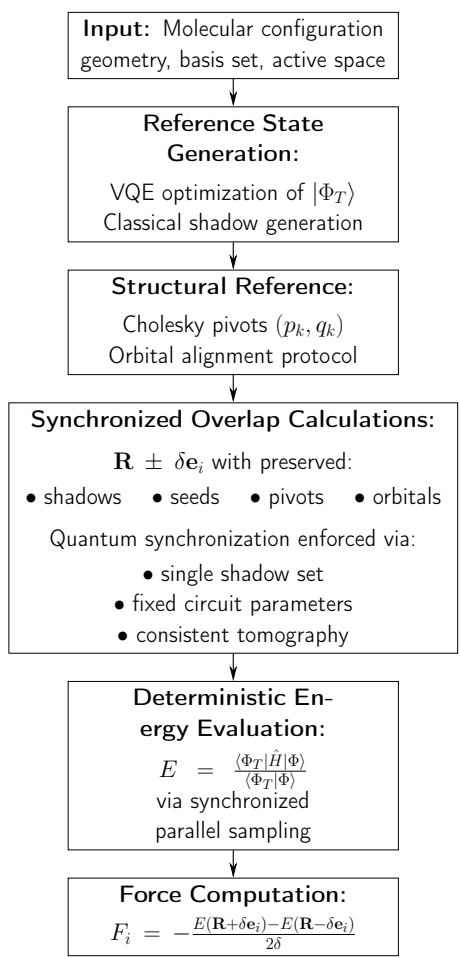

Figure \ref{fig:correlated-sampling} illustrates the complete correlated sampling workflow, highlighting the synchronization of quantum and classical random processes across displaced geometries.

\clearpage
\subsubsection{Deterministic Cholesky Decomposition}
\begin{figure}[htb]
    \centering
    \resizebox{0.85\textwidth}{!}{
        \begin{tikzpicture}[
    font=\sffamily\Large,
    node distance=1.35cm and 1.8cm,
    box/.style={
        rectangle, 
        draw,
        text width=7cm,
        minimum height=1.2cm,
        align=center,
        inner sep=5pt
    },
    decision/.style={
        diamond,
        draw,
        text width=3.5cm,
        inner sep=1pt,
        align=center,
        aspect=2
    },
    note/.style={
        rectangle,
        draw,
        dashed,
        text width=6cm,
        align=left,
        inner sep=5pt
    },
    arrow/.style={
        -{Stealth[length=4mm, width=2mm]},
        thick
    }
]
\node[box] (start) {
    \textbf{Input:} Molecule geometry\\ and electron integrals $V_{(pq)(rs)}$
};

\node[decision] (check) [below=0.9cm of start] {Reference state exists?};

\node[box, align=left] (load_ref) [below left=of check, xshift=-1.8cm, yshift=-0.9cm] {
    \textbf{Load reference state:}\\[0.2cm]
    1. Pivot sequence\\
    2. Reference orbitals\\
    3. Shell-to-AO mappings
};

\node[box, align=left] (gen_ref) [below right=of check, xshift=1.8cm, yshift=-0.9cm] {
    \textbf{Generate reference:}\\[0.2cm]
    1. Compute diagonal elements\\
    2. Select pivots by max residual\\
    3. Store and MPI broadcast
};

\node[box] (decomp) [below=3.6cm of check, yshift=-0.9cm] {
    \textbf{Deterministic decomposition:}\\[0.2cm]
    Follow stored pivot path $(p_k,q_k)$, else while $\delta > \epsilon$:\\[0.3cm]
    $L_{pq}^k = \dfrac{V_{(pq)(p_kq_k)} - \sum\limits_{\gamma} L_{pq}^{\gamma} L_{p_kq_k}^{\gamma}}{\sqrt{D_{p_kq_k}}}$
};

\node[decision] (validate) [below=2.7cm of decomp] {Decomposition valid?};

\node[box] (output) [below left=of validate, xshift=-1.35cm, yshift=-0.9cm] {
    Return Cholesky vectors $L_{pq}^{\gamma}$
};

\node[box] (reset) [below right=of validate, xshift=1.35cm, yshift=-0.9cm] {
    Reset reference state
};

\draw[arrow] (start.south) -- (check.north);
\draw[arrow] (check.south west) -- node[above left]{Yes} (load_ref.north);
\draw[arrow] (check.south east) -- node[above right]{No} (gen_ref.north);
\draw[arrow] (load_ref.south) |- ([xshift=-0.0cm]decomp.west);
\draw[arrow] (gen_ref.south) |- ([xshift=0.0cm]decomp.east);
\draw[arrow] (decomp.south) -- (validate.north);
\draw[arrow] (validate.south west) -- node[above left]{Yes} (output.north);
\draw[arrow] (validate.south east) -- node[above right]{No} (reset.north);
\draw[arrow] (reset.east) -- ++(1,0) coordinate (tmp)
  |- node[near start, left, yshift=1cm, xshift=-0.25cm]{Reset} (check.east);

\end{tikzpicture}
    }
    \caption{Flow diagram of the deterministic Cholesky decomposition algorithm. This approach ensures that the same pivots are selected for all geometrically displaced structures, maintaining strict correlation between QMC runs.}
    \label{fig:deterministic-cholesky}
\end{figure}
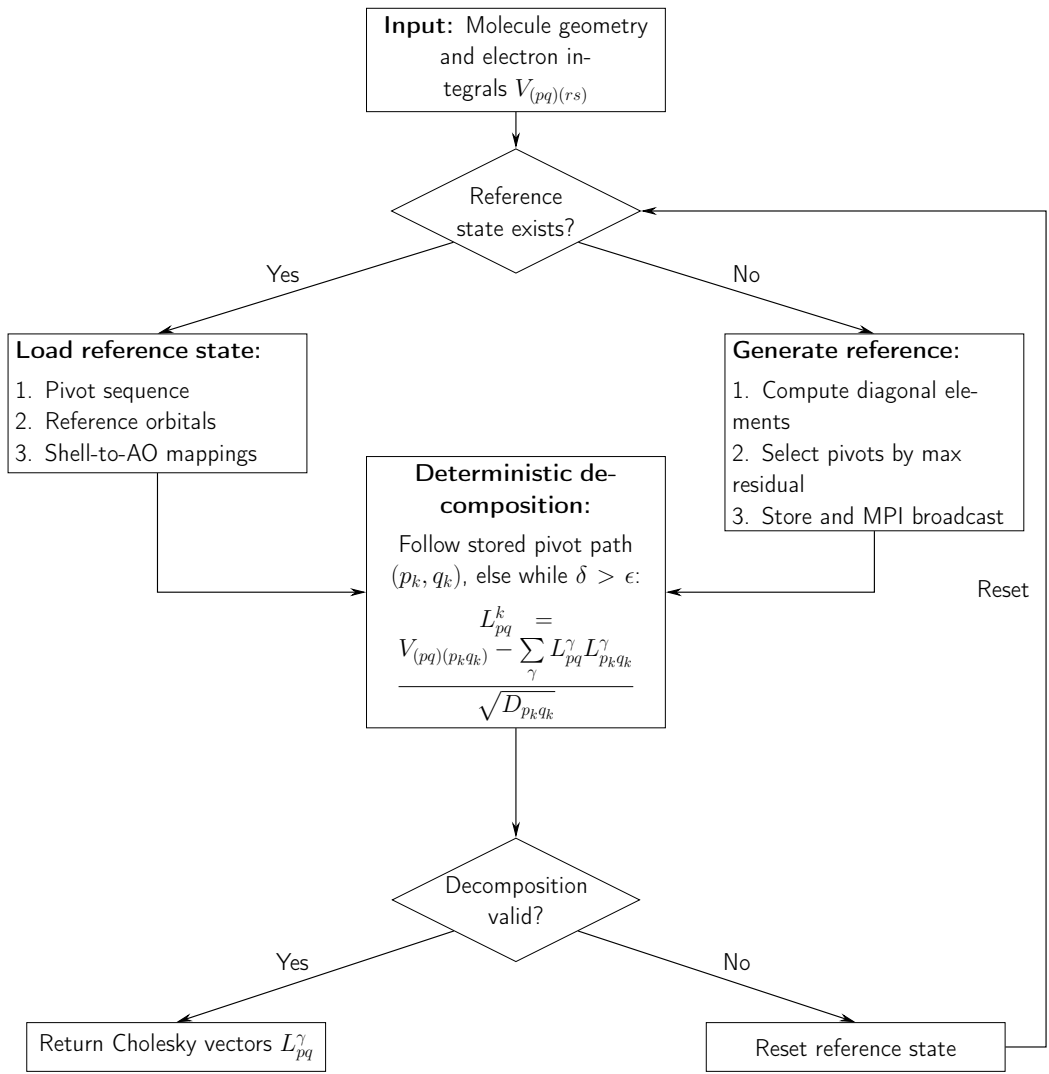

Figure \ref{fig:deterministic-cholesky} shows the algorithm for deterministic Cholesky decomposition, which ensures consistent pivot selection across geometric displacements.

\clearpage
\subsubsection{Molecular Orbital Alignment}
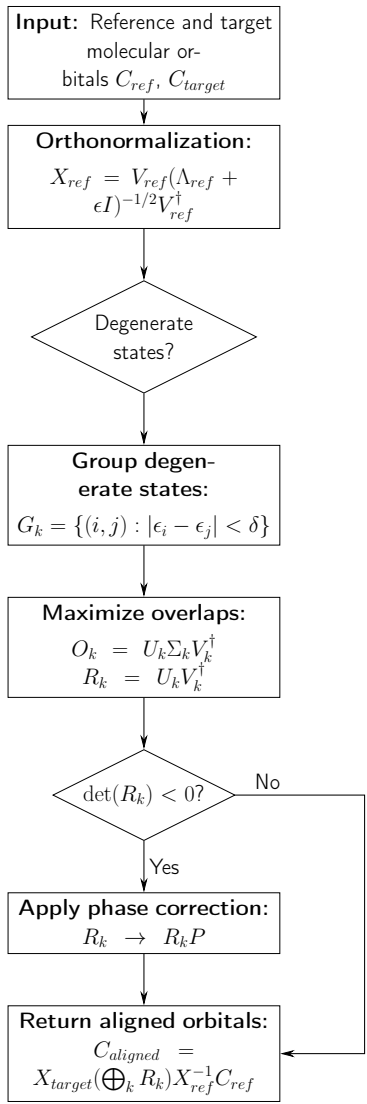
\begin{figure}[htb]
    \centering
    \resizebox{0.3\textwidth}{!}{
        \begin{tikzpicture}[
    font=\sffamily\Large,
    node distance=1.35cm and 1.8cm,
    box/.style={
        rectangle, 
        draw,
        text width=7cm,
        minimum height=1.2cm,
        align=center,
        inner sep=5pt
    },
    decision/.style={
        diamond,
        draw,
        text width=3.5cm,
        inner sep=1pt,
        align=center,
        aspect=2
    },
    note/.style={
        rectangle,
        draw,
        dashed,
        text width=6.5cm,
        align=left,
        inner sep=5pt
    },
    arrow/.style={
        -{Stealth[length=4mm, width=2mm]},
        thick
    }
]
\node[box] (start) {
    \textbf{Input:} Reference and target\\molecular orbitals $C_{ref}$, $C_{target}$
};
\node[box] (ortho) [below=0.7cm of start] {
    \textbf{Orthonormalization:}\\[0.2cm]
    $X_{ref} = V_{ref}(\Lambda_{ref} + \epsilon I)^{-1/2}V_{ref}^{\dagger}$
};
\node[decision] (check_degen) [below=1.4cm of ortho] {Degenerate states?};
\node[box] (handle_degen) [below=1.4cm of check_degen] {
    \textbf{Group degenerate states:}\\[0.2cm]
    $G_k = \{(i,j) : |\epsilon_i - \epsilon_j| < \delta\}$
};
\node[box] (svd) [below=1.4cm of handle_degen] {
    \textbf{Maximize overlaps:}\\[0.2cm]
    $O_k = U_k\Sigma_kV_k^{\dagger}$\\
    $R_k = U_kV_k^{\dagger}$
};
\node[decision] (check_phase) [below=1.4cm of svd] {$\det(R_k) < 0$?};
\node[box] (phase) [below=1.4cm of check_phase] {
    \textbf{Apply phase correction:}\\
    $R_k \rightarrow R_k P$
};
\node[box] (output) [below=1.4cm of phase] {
    \textbf{Return aligned orbitals:}\\
    $C_{aligned} = X_{target}(\bigoplus_k R_k)X_{ref}^{-1}C_{ref}$
};
\draw[arrow] (start.south) -- (ortho.north);
\draw[arrow] (ortho.south) -- (check_degen.north);
\draw[arrow] (check_degen.south) -- (handle_degen.north);
\draw[arrow] (handle_degen.south) -- (svd.north);
\draw[arrow] (svd.south) -- (check_phase.north);
\draw[arrow] (check_phase.south) -- node[right]{Yes} (phase.north);
\draw[arrow] (phase.south) -- (output.north);
\draw[arrow] (check_phase.east) -- node[above, near start]{No} ++(3.5,0) |- (output.east);

\end{tikzpicture}
    }
    \caption{Schematic overview of the molecular orbital alignment procedure. This algorithm ensures that orbitals maintain consistent character across geometrically displaced structures, handling degeneracies and phase corrections.}
    \label{fig:mo-alignment}
\end{figure}

Figure \ref{fig:mo-alignment} depicts the orbital alignment procedure that maintains consistent orbital descriptions across geometric displacements, handling orbital degeneracies and phase consistency.

\bibliography{references} 

\end{document}